\documentclass[12pt]{elsarticle} 

\usepackage{float}                            
\usepackage[utf8]{inputenc}
\usepackage{amsmath}
\usepackage{amssymb}
\usepackage{amsbsy}
\usepackage{graphicx}
\usepackage{algorithm,algorithmic}
\usepackage{tabularx}
\usepackage{siunitx}
\usepackage{natbib}
\usepackage{subfig}
\usepackage{setspace}
\usepackage[left=3.5cm,right=2.5cm,top=2.5cm,bottom=2cm,includeheadfoot]{geometry}

\makeatletter
\def\ps@pprintTitle{%
 \let\@oddhead\@empty
 \let\@evenhead\@empty
 \def\@oddfoot{}%
 \let\@evenfoot\@oddfoot}
\makeatother

\newcommand{\Var}[1]{\operatorname{Var}(#1)}
\newcommand{\E}[1]{\operatorname{E}(#1)}

\begin{document}

\begin{frontmatter}
\title{A Bayesian Time-Varying Autoregressive Model for Improved Short- and Long-Term Prediction}
\author[CB]{Christoph Berninger}
\author[AS]{Almond St\"{o}cker}
\author[CB]{David R\"{u}gamer}
\address[CB]{Department of Statistics, LMU München}
\address[AS]{School of Business and Economics, Humboldt University of Berlin}
\begin{keyword}MCMC Metropolis-Hastings \sep%
    Gibbs sampler \sep%
    Bayesian time-varying autoregressive models \sep%
    long run regluarization \sep%
    interest rate models
\end{keyword}
\begin{abstract}
Motivated by the application to German interest rates, we propose a time-varying autoregressive model for short and long term prediction of time series that exhibit a temporary non-stationary behavior but are assumed to mean revert in the long run. We use a Bayesian formulation to incorporate prior assumptions on the mean reverting process in the model and thereby regularize predictions in the far future. We use MCMC-based inference by deriving relevant full conditional distributions and employ a Metropolis-Hastings within Gibbs sampler approach to sample from the posterior (predictive) distribution. In combining data-driven short term predictions with long term distribution assumptions our model is competitive to the existing methods in the short horizon while yielding reasonable predictions in the long run. We apply our model to interest rate data and contrast the forecasting performance to the one of a 2-Additive-Factor Gaussian model as well as to the predictions of a dynamic Nelson-Siegel model.
\end{abstract}
\end{frontmatter}

\newpage

\section{Introduction}
To forecast an univariate time series the first model of choice is often a linear model. An archetype of this model class in the context of time series analysis is the autoregressive model of order 1 (AR(1)), which is defined as follows:
\begin{align}
	\label{eq:AROne}
	x_t = \alpha + \beta x_{t-1} + \epsilon_t,
\end{align}
where $x_t$ represents the observed variable at time point $t$ and $\alpha$ and $\beta$ are real valued constants, while $|\beta| < 1$ is assumed to ensure stationarity. The innovation process $\epsilon_t$ can be, e.g., a Gaussian white noise process, i.e., an independent and identically (i.i.d.) normal distributed  $\epsilon_t \overset{i.i.d.}{\sim} \mathcal{N}(0,\sigma^2)$ for all time points $t \in \mathcal{T}$. In this classical model, the one-step-ahead expectation $\E{x_t\mid x_{t-1}} = \alpha + \beta x_{t-1}$ and variance $\Var{x_t\mid x_{t-1}} = \sigma^2$ are closely linked to the marginal characteristics $\E{x_t} = \frac{\alpha}{1-\beta}$ and $\Var{x_t} = \frac{\sigma^2}{1-\beta^2}$ approached in the long run. In this sense, fitting the short-term behavior of a time series with a linear model has wide implications for its long-term behavior, and, conversely, controlling the long-term behavior of the model constraints its short-time fit.
In practice, this close linkage may present an important limitation when short-term performance conflicts with long-term plausibility. 
\\
This gets evident %, i.a., 
in the macroeconomic literature and, more specifically, in contrasting modeling approaches for interest rates, which motivate this work: Among others, \citet{diebold2006} focus on predictions in the short horizon. The authors apply an AR(1) process to extracted factors of the yield curve. While they do make the long-term assumption that interest rates are principally mean reverting, the process exhibits an almost integrated behavior. Estimating the model parameters of a near integrated but stationary AR(1) model might give large estimation errors and lead to unrealistic long run mean estimates far beyond the range of the data. \citet{Duffee2011} even discard the mean reversion / stationarity assumption. In their proposed random walk model, the prediction variance linearly increases in time leading to extreme values in the long run. The strong focus on the short horizon might lead to questionable and potentially implausible long-term behavior. In contrary, \citet{KornWagner2020} apply a linear model to model long horizon features. Their proposal is a Gauss2++ model with a forward looking estimation approach, i.e., they calibrate the parameters of their model without using historical data but amongst current prices of interest rate derivatives and long run survey forecasts. The Gauss2++ model is a standard model in the insurance industry, where plausible forecasts in the longer run are required.
As demonstrated later in this paper, their model achieves a realistic long run distribution, however, at the expense of inferior short term predictions.
\\
\\
The aim of this paper is to bridge the gab between short and long horizons, generalizing above mentioned approaches to a model with the flexibility to a) sufficiently adapt to sample data to yield good short-term predictions and b) apply suitable regularization to achieve plausible long- and middle-term forecasts at the same time. This is particularly, yet not exclusively, relevant in applications where a stable stationary distribution is assumed in the long run -- despite observing a 'temporary non-stationary behavior' in the available data where, e.g., unconstrained linear model fits or a Dickey-Fuller test \citep{DickeyFuller1979} would indicate an integrated process.
\\
\\
As in general linearity is often a restrictive assumption in practice and many time series exhibit features that cannot be captured by a linear model (\citet{Hamilton1989}), a lot of research has been conducted to introduce different types of nonlinear models in the last decades. In particular, nonlinear models offer more flexibility to account for both, short and long horizon.
\\
A bi-linear model is an example of a nonlinear model, which assumes a nonlinear relationship between the covariates and response variable (see, e.g., \citet{Granger1978, Rao1984}), although not often used in macroeconomic applications (\citet{Morley2009}). A more immediate approach is to allow one (or more) parameters of a linear model to change over time. This comprises the regime switching and time-varying parameter models.
\\
\\
Regime switching models can allow for a different mean reversion level in the short and the long horizon. This feature can be used to regularize the long run mean of the model. The first approaches to regime switching models were conducted by \citet{Quandt1958}, who considered a switching regression model extending a linear regression model by allowing the parameters to switch between different states according to a random variable. \citet{Bacon1971} introduced a smooth transition model, which implements a smooth transition from one regime to another without a sudden jump. \citet{Goldfeld1973} introduced the Markov switching regression model and use a discrete latent Markov process to determine the current regime. These models were adapted to time series models by \citet{Tong1980} and \citet{Chan1986} introducing the threshold autoregressive model (TAR) and the smooth transition autoregressive model (STAR), respectively. \citet{Hamilton1989} introduced the Markov switching autoregressive model for applications in economics, which have been investigated thoroughly together with different variants in the literature (see, e.g.,  \citet{Haggan1981, Jansen1996, Terasvirta1994}). \citet{Lanne2002} used a TAR-model, which only allows regime changes for the constant parameter $\alpha$, and applied it to strongly autocorrelated time series data -- which is very related to temporary non-stationary behavior of the time series and, therefore, to the aim of the paper. When there is, however, no concrete indication for the process dynamics to result from switches between discrete underlying states, we consider it more natural and promising to assume a continuous latent process.
\\
\\
In contrast to regime switching models, time-varying parameter models allow one (or more) of the parameters in a linear model to be driven by its own continuous process (\citet{Morley2009}). For example, if the parameter vector ($\alpha$, $\beta$, $\sigma^2$) of the linear AR(1) model becomes a stochastic process, this results in a time-varying autoregressive model of order 1 (TV-AR(1))
\begin{align}
	x_t = \alpha_t + \beta_t x_{t-1} + \epsilon_t
\end{align}
with $\epsilon_t \sim \mathcal{N}(0,\sigma_t^2)$. Certain distribution assumptions for the underlying stochastic process of the parameter vector ($\alpha_t$, $\beta_t$, $\sigma_t$) are made in practice to complete the TV-AR(1) model specification (\citet{Terasvirta2010}).
Similar to the TAR model in \citet{Lanne2002} the time variation of the TV-AR(1) model can be restricted to the constant parameter $\alpha_t$, resulting in a time-varying constant autoregressive model of order 1 (TVC-AR(1)):
\begin{align}
	x_t = \alpha_t + \beta x_{t-1} + \epsilon_t.
\end{align}
If $|\beta| < 1$ and the latent process of $\alpha_t$ is stationary, the process $x$ is also stationary. But due to random shifts in the mean reversion level -- because of the time-varying constant parameter -- realizations of the model can resemble those of a (close to) random walk process, when restricting to a limited time window.
\\
Another time-varying parameter model is the shifting endpoint model introduced by \citet{Kozicki2001}. Similar to the TVC-AR(1) model their approach has a time-varying mean reversion level, referred to as shifting endpoints. In particular, \citet{Dijk2014} applied this model to interest rates presenting a slow moving trend using exponential smoothing or long survey forecasts. There is also a strand of literature, which connects the level of interest rates in the long run to the expected inflation dynamics, also referred to as trend inflation (see, e.g., \citet{bekaert2010}, \citet{cieslak2015}, \citet{Kozicki2001}, \citet{rudebusch2008}). Associating the variable of interest with appropriate covariates might practically help in several scenarios, but does not directly address the core of the present problem.
\\
\\
In this paper, we propose a model approach competitive in terms of short horizon forecasts, yet controlled to obtain realistic predictive distributions for the long horizon. We propose a Bayesian TVC-AR(1) model, which is stationary but can resemble short-term properties of an integrated or nearly integrated linear process due to a stochastic mean reversion level. The model allows us to regularize the long run distribution of the time series without affecting short term distributions adversely. Different to \citet{Dijk2014} we do not use a deterministic shifting mean reversion level, but incorporate long run assumptions via prior information in a Bayesian approach, such that the latent coefficient process, and with it the mean reversion dynamics, are still estimated from the data. \\ 
In particular, the novelty of our approach lies in the model allowing us to (1) regularize long run predictions by using prior assumptions, (2) separate the modeling process into a data driven short horizon model-part and a long horizon model-part that is determined by prior (or expert) assumptions and (3) yield improved forecasting performance in the short horizon compared to the commonly used linear dynamic Nelson-Siegel model and Gauss2++ model, while retaining realistic long run distributions. Moreover, we place particular emphasis on the interpretability of the model structure and prior parameters, preserving a close link to the common linear models. This allows to include expert knowledge or assumptions in accordance with economic theory about the long run behavior of a time series into the model in a sound mathematical way. We here specifically focus on the application to interest rates. As our model allows to regularize long run predictions, it is also of particular interest for insurance companies, where realistic long run interest rate forecasts are needed to evaluate the risk and performance of specific insurance products.
\\
\\
The remainder of this paper is arranged as follows. Section \ref{sec:Theory} specifies the Bayesian TVC-AR(1), including the derivation of required full conditional posterior distributions and the application of a Metropolis-Hastings within Gibbs sampling routine for statistical inference. In Section \ref{sec:Application} we discuss an application of our model to interest rate data and compare the forecasting performance as well as the long run distribution of our nonlinear model with the dynamic Nelson-Siegel model (short-term focus) and the Gauss2++ model (long-term focus). We conclude with Section \ref{sec:Conclusion} and give a brief outlook on potential further research topics.

\section{A Bayesian TVC-AR(1) Model for Long Run Regularization}
\label{sec:Theory}
In this Section we introduce the Bayesian TVC-AR(1) (BTVC-AR(1)) model. The model incorporates assumptions about the long-term behavior of the time series and thereby regularizes the process in the long horizon. At the same time, the model is mainly driven by the given data in the short run and thus fosters a good short-term prediction.

\subsection{The BTVC-AR(1) Model} 
\label{sec:modelSpec}
The BTVC-AR(1) model is defined as follows:
\begin{align}
\label{eq:TVCAR}
x_t = \alpha_t + \beta x_{t-1} + \epsilon_t,
\end{align}
where $\beta$ represents the mean reversion speed and $|\beta|<1$ to secure stationarity. $\epsilon_t$ is assumed to be a Gaussian white noise process, i.e., $\epsilon_t \overset{i.i.d.}{\sim} \mathcal{N}(0,\sigma^2)$. We further specify $\alpha_t$ as a stationary Gaussian process specified by its unconditional expectation $\boldsymbol{\theta} := \vartheta \cdot\boldsymbol{1}$ and covariance $\boldsymbol{\Sigma}$, i.e.,
\begin{align} \label{eq:alpha}
\boldsymbol{\alpha} := (\alpha_1, \alpha_2, ... , \alpha_t) \sim \mathcal{N}_t(\boldsymbol{\theta}, \boldsymbol{\Sigma}).
\end{align}
The Bayesian approach considers the parameters of model (\ref{eq:TVCAR}) as random variables. For the conditional prior distribution of $\beta$ conditional on  $\sigma^2$ a truncated normal distribution with lower bound $-1$ and upper bound $1$ is assumed as a prior, i.e.,
\begin{align*}
\beta | \sigma^2 \sim \mathcal{N}(\mu_{\beta}, \sigma^2 \cdot \sigma_{\beta}^2, -1, 1),
%\beta | \sigma^2 \sim \mathcal{N}(\mu_{\beta}, \sigma^2\sigma_{\beta}^2, l(\sigma^2), u(\sigma^2)).
\end{align*}
with conditional prior expectation $\mu_\beta$ and additional multiplicative variance parameter $\sigma_\beta$.
The prior distribution for $\sigma^2$ is an inverse gamma distribution with shape and scale parameter, $a$ and $b$, respectively,
\begin{align*}
\sigma^2 &\sim \mathcal{IG}(a, b).
\end{align*}
These two prior distributions are conjugate priors for model (\ref{eq:TVCAR}) if the respective other parameter is known and therefore allow for an analytical derivation of the corresponding full conditional distributions.

Using these priors the defined model can be seen as a Bayesian version of the TVC-AR(1) model. The mean $\boldsymbol{\theta}$ and covariance $\boldsymbol{\Sigma}$ might be assumed fixed or defined as random variables with further attached prior distributions. In the latter case (\ref{eq:alpha}) describes the distribution of $\boldsymbol{\alpha}$ conditional on $\boldsymbol{\theta}$ and $\boldsymbol{\Sigma}$. Placing priors on these parameters allows to incorporate assumptions about the long run distribution into the model as further elaborated in Section \ref{sec:ArbitratingShortandLong}.
\\

While this basic model setup is flexible in many ways and particular in terms of its covariance structure assumptions for $\boldsymbol{\alpha}$, further practical insights can be obtained from a more in-depth model characterization. In the following, we will shed light on useful properties of this framework when assuming an AR-covariance structure. This covariance structure has shown to provide good results in applications.

\subsection{Arbitrating Between Short and Long Run Distribution}
\label{sec:ArbitratingShortandLong}
The goal of our work is to propose a new modeling framework, which can regularize the long run distribution of (nearly) integrated time series by keeping a good forecasting performance in the short horizon. Linear models often concentrate on the conditional distribution in the short horizon, but due to the near integration property of the time series this can lead to inappropriate long run distributions. For example, if the AR(1) model is estimated for a time series, which shows a (close to) random walk behavior, the parameter $\beta$ of the model will take a value close to $1$. This can lead to a large long run variance given by
$$
\frac{\sigma^2}{1-\beta^2},
$$
potentially yielding unrealistic values in the long run that have never been observed in the past. On the other hand, calibrating $\beta$ to a given long run variance is not straightforward without deteriorating the short run prediction performance. Figure \ref{fig:ModelComparison} depicts this undesired behavior by showing the long run mean of a linear AR(1) model which is driven by the conditional short run distribution at the expense of an unrealistic long-term distribution. \\

We address this issue by incorporating a time-varying mean reversion level, which locally preserves the good short term prediction and at the same time regularizes the long run distribution. The current mean reversion level valid in the short run can be different to the long run behavior accounting for the current market situation and therefore improving the short run prediction. We enable the model to stay in a reasonable range in the long run by assuming a stationary process for the time-varying mean reversion level and a stronger mean reversion to this time-varying level than a linear AR(1) model would induce to its constant mean reversion level. Such a behavior can be achieved by introducing a time-varying $\alpha$ parameter into a linear AR(1) model with additional prior assumptions. In particular, this does not change the (weak) stationarity property of the model if the assumed process for $\alpha$ is (weakly) stationary itself. This can be verified by calculating the unconditional mean, the unconditional variance and the unconditional covariance: 

\begin{align*}
    E[x_t] &= \frac{\vartheta}{1- \beta} \\
    Var[x_t] &= \frac{\sigma^2}{(1-\beta^2)} + \frac{Var(\alpha_t) + 2\beta Cov(\alpha_t,x_{t-1})}{(1-\beta^2)} \\
    Cov(x_{t-h}, x_t) &= \sum_{i=0}^{h} \beta^i Cov(\alpha_{t-i}, x_{t-h}) + \beta^h Var(x_{t-h})
\end{align*}
As the $\alpha$-process is stationary and 
$$
Cov(\alpha_t, x_{t-h}) = \sum_{i=0}^\infty \beta^i Cov(\alpha_t, \alpha_{t-h-i}),
$$
the BTVC-AR(1) is (weakly) stationary.

The time-varying $\alpha$ increases the flexibility of our model to account for short and long run distributional properties. As current observations have almost no influence in the very long run, a reasonable way to include information about the long run mean and long run variance in a Bayesian setting is via prior assumptions for $\boldsymbol{\alpha}$. We will further elaborate this in Section \ref{sec:LongRunMean} and \ref{sec:LongRunVariance}.
The time-varying $\alpha$ also increases the flexibility of the model such that the conditional distribution in the short run is consistent with the empirical data, i.e., $E[x_{t+h}|x_t, x_{t-1},...]$ and $Var[x_{t+h}| x_t, x_{t-1},...]$ still reflect the empirical distribution for a short horizon $h$.

Our BTVC-AR(1) model can therefore produce both, a conditional short term distribution, which roughly corresponds to an unrestricted linear model, and a long run distribution with a reasonable range of values.

\subsubsection{The Long Run Mean and Time-Varying Mean Reversion}
\label{sec:LongRunMean}
The mean reversion level in a linear AR(1) model as specified in (\ref{eq:AROne}) amounts to 
$$
\frac{\alpha}{1-\beta}.
$$
As the mean reversion level stays constant over time it is also the long run mean of the model. In contrast, the mean reversion level in the BTVC-AR(1) model changes over time and is given by
$$
\frac{\alpha_t}{1-\beta}
$$
for time point $t$. This local mean reversion level is in general different to the long-term mean and can even pull the process away from it in expectation, i.e,
$$
\big|\,E[x_{t+h}|x_t, x_{t-1},...] - E[x_t] \,\big| \geq \big| \, x_{t} - E[x_t] \,\big|,
$$
which helps fitting the model to a time series exhibiting a (close to) random walk behavior. The long run mean of the BTVC-AR(1) depends on the unconditional mean of $\boldsymbol{\alpha}$ and amounts to
$$
\frac{\vartheta}{1- \beta}
$$
in our model. We assume the data to be centered around a prior specified long run mean. By setting $\boldsymbol{\theta} = \boldsymbol{0}$, i.e., $\vartheta = 0$, this long run mean is reached in expectation after reshifting the simulated data. 
\\

\begin{figure}
  \centering
  \subfloat[][]{\includegraphics[width=0.52\linewidth]{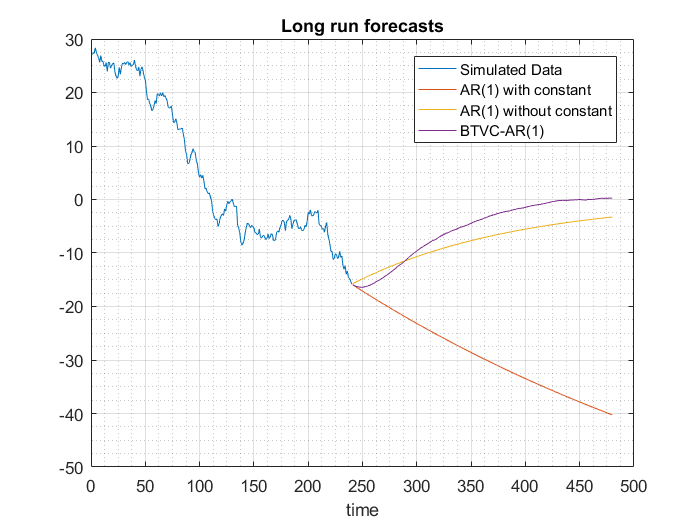}}%
  \subfloat[][]{\includegraphics[width=0.52\linewidth]{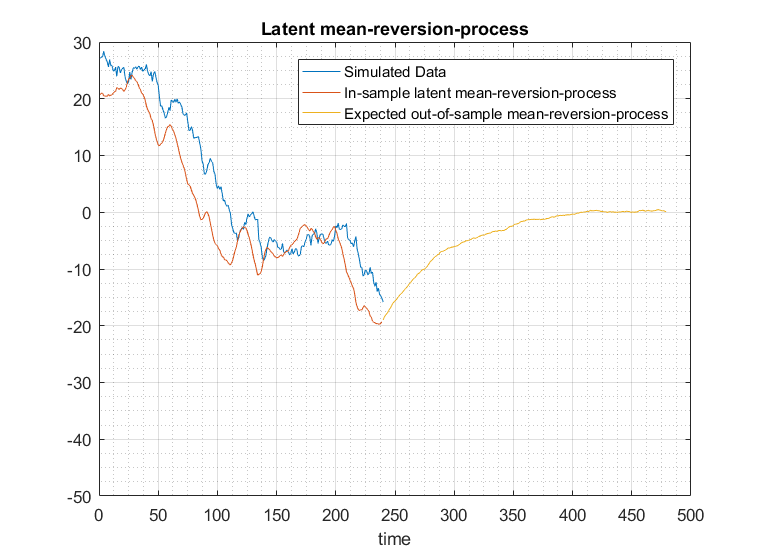}}%
  \caption{A comparison of a linear AR(1) model with no restrictions for the constant parameter, a linear AR(1) model with restrictions to the constant parameter and a BTVC-AR(1) model applied on a simulated time series.}%
  \label{fig:ModelComparison}
\end{figure}

The implications of the time-varying mean reversion level of the BTVC-AR(1) model are visualized in Figure \ref{fig:ModelComparison}. Two AR(1) models (with unrestricted and restricted constant parameter) and the BTVC-AR(1) model have been exemplary fitted to a simulated stationary time series, which shows a (nearly) integrated behavior. 

In the left graphic the ``historical'' time series can be seen as well as the expected future development according to the three models. The AR(1) model with no restrictions has a long-term mean far away from the historical domain, as its focus lies on the conditional short term distribution. The restricted AR(1) model sets the $\alpha$ parameter to $0$ to regularize the long run mean, but at the same time the expected values in the short horizon are pushed in the direction of the long run level leading to an inferior forecasting performance. If we assume that the (close to) random walk behavior stems from changes in the mean reversion level determined by unobserved variables, the BTVC-AR(1) model has a more desired behavior. The time-varying constant parameter in the model leads to a time-varying mean reversion level and can therefore account for the changes induced by the unobserved variables. The long run mean can still be regularized to $0$ while influencing the short term distribution less abruptly. This allows the time series to follow the current trend in expectation and veer away from the long run mean for a couple of time steps. The reason for this behavior is that the latent $\alpha$-process induces a local mean reversion level that lies below the last observation, which can be seen in the right plot of Figure \ref{fig:ModelComparison} showing the average latent mean reversion level extracted during the simulation process. In the long run the mean reversion level returns in expectation to the prespecified value of $0$.

\subsubsection{Long Run Variance}
\label{sec:LongRunVariance}
The long run variance of a linear AR(1) model is given by
$$
\frac{\sigma^2}{1-\beta^2}.
$$
The closer the model behaves like a random walk, i.e., the closer $\beta$ approaches $1$, the larger the long run variance gets under the assumption of a fixed conditional variance $\sigma^2$. In terms of the long run variance, the BTVC-AR(1) model is more flexible by incorporating two sources of variation, the residual term of the AR(1) model and variance of the latent $\alpha$-process. The model's long run variance is given by
\begin{align}
    Var(x_t) = \frac{\sigma^2}{1-\beta^2} + \frac{Var(\alpha_t) + 2\beta Cov(\alpha_t,x_{t-1})}{1-\beta^2}.
    \label{eq:uncondVariance}
\end{align}
The first term has the same form as the long run variance of a linear AR(1) model and can be interpreted as the ``unconditional'' variance around the time-varying mean reversion level, i.e., the variance conditional on the $\alpha$-process. The second term incorporates the part of the variance stemming from the $\alpha$-process and depends on both its unconditional variance and unconditional covariances. This allows the BTVC-AR(1) model to be more flexible and to control the long run variance of $x_t$, while reducing the opposing effect on the conditional distribution in the short horizon. The model thus still produces short term distributions consistent with the given data. If $\alpha$ is a constant process, the second term is $0$ and the BTVC-AR(1) model reduces to a linear AR(1) model.
\\

\paragraph{Prior Assumptions} With the goal in mind to control the long-term variance based on prior information, a more refined specification of the BTVC-AR(1) model is helpful in order to translate this information into the model. We will use a centered $\alpha$-process with an AR-covariance structure for demonstrative purposes. In this case, $\boldsymbol{\alpha}$ can be represented by a linear AR(1) model 
$$
\alpha_t = \rho \alpha_{t-1} + \eta_t,
$$
where $\rho$ represents the correlation between two successive time steps and $\eta_t$ is an i.i.d. Gaussian white noise process, i.e., $\eta_t \overset{i.i.d.}{\sim} \mathcal{N}(0,\tau^2)$.
The long run variance of the BTVC-AR(1) model is then given by
\begin{align}
    Var(x_t) = \frac{\sigma^2}{(1-\beta^2)} + \frac{\tau^2(1+\rho\beta)}{(1-\rho\beta)(1-\beta^2)(1-\rho^2)}.
    \label{eq:uncondVarianceAR}
\end{align}
If the process $x_t$ is supposed to reach a certain objective variance in the long run, the degrees-of-freedom in (\ref{eq:uncondVarianceAR}) reduce from four to three. For example, for given $\rho$, $\beta$ and $\sigma^2$ and a prior value assumption for $Var(x_t)$, the variance of $x_t$ has a one-to-one relationship with $\tau^2$ and it is straightforward to solve (\ref{eq:uncondVarianceAR}) for $\tau^2$. Let denote the solution by $\tilde{\tau}^2$. To ensure positivity the truncation limits for the prior distribution of $\beta$ can be set to $-1$ and $\sqrt{\frac{Var(x_t) - \sigma^2}{Var(x_t)}}$. For this specific covariance structure, a possible prior distribution of $\tau^2$ can thus be defined by the conditional distribution
\begin{align}
    \label{eq:priorTauSq}
    \tau^2 | \rho, \beta, \sigma^2 \sim
    \delta_{\tilde{\tau}^2},
\end{align}
where $\delta$ denotes a degenerated distribution with point mass 1 at $\tilde{\tau}^2$. This definition forces the process to reach its prespecified long run variance ${Var}(x_t)$ while controlling the speed of mean reversion of the $\alpha$-process through $\rho$. A conjugate prior for $\rho$ is a normal distribution truncated below by $-1$ and from above by $1$, i.e., 
\begin{align}
    \label{eq:priorRho}
    \rho \sim \mathcal{N}(\mu_{\rho}, \sigma^2_{\rho}, -1, 1),
\end{align}
with mean $\mu_\rho$ and variance $\sigma^2_\rho$.

The previous prior specifications allow to introduce prior information into the model in a straightforward manner while maintaining the properties of the BTVC-AR(1) model.

\subsubsection{The Short Run Distribution}

For the short run distribution of the BTVC-AR(1) model the goal is to balance between a consistent estimation with the observed data and the opposing effect of the prespecified long run distribution. For a linear AR(1) model with a restricted long run mean of $0$ the conditional expectation and the conditional variance amount to 
\begin{align*}
    E[x_{t+1}|x_t,...] &= \beta x_t, \\
    Var[x_{t+1}|x_t,...] &= \sigma^2.
\end{align*}
The model can get arbitrarily close to a centered random walk if $\beta$ approaches $1$, while the long run variance increases at the same time as shown in Section \ref{sec:LongRunVariance}. For the BTVC-AR(1) model we get 
\begin{align*}
    E[x_{t+1}|x_t,...] &= E[\alpha_{t+1}|x_t,...] + \beta x_t, \\
    Var[x_{t+1}|x_t,...] &= Var[\alpha_{t+1}|x_t,...] + \sigma^2.
\end{align*}
A random walk behavior, i.e., $E[x_{t+1}|x_t] \approx x_t$, can be reached without $\beta$ necessarily being close to $1$ due to the conditional expectation of the $\alpha$-process that supports the random walk behavior in the short horizon. This increases the flexibility of the BTVC-AR(1) model compared to a linear AR(1) model in combining short and long run distributional characteristics.
\\

We can further decompose the conditional expectation to see the similarities of the BTVC-AR(1) model to a linear AR(t) process at a given time point $t$. Let \linebreak ${\boldsymbol{\breve{\alpha}} = (\alpha_1, ..., \alpha_{t+1})}$ denote the time-varying constant extended to $t+1$ in a consistent manner with the BTVC-AR(1) model definition, i.e., the same covariance parameterization is assumed. For a given data set $\boldsymbol{x} = (x_0, ... , x_t)$, the conditional distribution of $\boldsymbol{\breve{\alpha}}|\boldsymbol{x}$ is multivariate normal (c.f. \ref{appendix:FC_alpha}), i.e.,
$$
\boldsymbol{\breve{\alpha}} | \boldsymbol{x} \sim \mathcal{N}(\boldsymbol{\breve{\mu}}, \boldsymbol{\breve{\Sigma}}).
$$
As $\boldsymbol{\breve{\mu}} = \frac{1}{\sigma^2}\boldsymbol{\breve{\Sigma}}\boldsymbol{\tilde{\Delta}}$ with  $\boldsymbol{\tilde{\Delta}} = (x_1- \beta x_0,...,x_t - \beta x_{t-1}, 0)^\top$, the conditional expectation of $\alpha_{t+1}$ is given by the last entry of $\boldsymbol{\breve{\mu}}$, 
$$
E[\alpha_{t+1}|\boldsymbol{x}] = \frac{1}{\sigma^2}\boldsymbol{s_{t+1,.}}\boldsymbol{\tilde{\Delta}},
$$
where $\boldsymbol{s_{t+1,.}} = (s_{t+1,1}, \ldots, s_{t+1,t+1})$ and $s_{i,j}$ represent the entries of $\boldsymbol{\breve{\Sigma}}$. The one step ahead conditional expectation of the model therefore amounts to
$$
E[x_{t+1}|x_t,...] = \left(\frac{s_{t+1,t}}{\sigma^2} + \beta\right)x_t + \sum_{i=1}^{t-1} \frac{s_{t+1,t-i} - \beta s_{t+1,t-(i-1)}}{\sigma^2}x_{t-i} - \frac{s_{t+1,1}}{\sigma^2}x_0.
$$
This shows that the conditional expectation depends on all previous time points like in a linear AR(t) model, allowing the BTVC-AR(1) model to better account for current trends in the process. Due to the given covariance structure for $\boldsymbol{\alpha}$ the number of parameters are, however, much less than in an actual AR(t) process.

\subsection{Bayesian Inference}
\label{sec:BayesianInference}
The main parameters of interest in the BTVC-AR(1) model are $\boldsymbol{\widetilde{\alpha}}$, $\beta$ and $\sigma^2$ with $\boldsymbol{\widetilde{\alpha}}$ extending $\boldsymbol{\alpha}$ by future time points up to the modeling horizon $h$, i.e.,
$$
\boldsymbol{\widetilde{\alpha}} = (\alpha_1,...,\alpha_t,...,\alpha_{t+h}).
$$
This extension is necessary to sample from the predictive posterior distribution of the parameters and to generate forecasts. The prior distribution of $\boldsymbol{\widetilde{\alpha}}$ incorporates the same assumptions as the prior distribution of $\boldsymbol{\alpha}$, i.e., 
\begin{align*}
    \boldsymbol{\widetilde{\alpha}} \sim \mathcal{N}_{t+h}(\boldsymbol{\widetilde{\theta}},\boldsymbol{\tilde{\Sigma}}) 
\end{align*}
where
\begin{equation*} 
\boldsymbol{\widetilde{\theta}} = \vartheta\cdot\boldsymbol{1} \quad \text{and} \quad \boldsymbol{\tilde{\Sigma}} = \begin{pmatrix}  \boldsymbol{\Sigma} & \boldsymbol{\Pi_{t+1}} & \dots & \boldsymbol{\Pi_{t+h}} \\ \boldsymbol{\Pi_{t+1}^\top} & \sigma_{\alpha}^2 \\
\vdots & &\ddots&\\
\boldsymbol{\Pi_{t+h}^\top}& & & \sigma_{\alpha}^2\end{pmatrix}
\end{equation*}
with $\boldsymbol{\Pi_{t+j}} = \{\text{Cov}(\alpha_{t+j},\alpha_1), \ldots, \text{Cov}(\alpha_{t+j},\alpha_{t+j-1})\}^\top$, i.e., the vector of covariances of $\alpha_{t+j}$ and all previous time points $1,\ldots,(t+j-1)$. For these time points the same (autoregressive) covariance parameterization as for $\boldsymbol{\alpha}$ is assumed for consistency reasons. $\sigma_{\alpha}^2$ represents the unconditional variance of the latent $\alpha$-process.

The goal of Bayesian inference is to find the joint posterior distribution, $p(\boldsymbol{\widetilde{\alpha}}, \beta, \sigma^2 | \boldsymbol{x})$, conditional on the observed data $\boldsymbol{x} = (x_0, ..., x_t)$.
If the full conditional distribution of all parameters is known, the Gibbs sampler (see, e.g. \citet{gelman2013bayesian}) can be used to draw samples from this joint posterior distribution and inference can be based on Monte Carlo approximation (see, e.g., \citet{Chib2001}). By regularizing the long run variance under the assumption of an AR-covariance structure and choosing a degenerated prior distribution for $\tau^2$ as in (\ref{eq:priorTauSq}), the full conditional distributions of $\rho, \beta$ and $\sigma^2$ depend on the prior of $\tau^2$ and can not be derived analytically. We therefore apply a Metropolis-Hastings within Gibbs sampling routine (see, e.g., \citet{Millar2000}). We will state the algorithmic details in the following section and here only derive the necessary distributions.\\

As the model defined in Section \ref{sec:LongRunVariance} can be considered under a different parameterization where $\tau^2$ is given by the function
$$
\tau^2 = f(\rho, \beta, \sigma^2, Var(x_t)) 
$$
and thus fixed for given $\rho, \beta, \sigma^2$ and a specified long run variance $Var(x_t)$, we will focus on deriving two conditional distributions in order to be able to employ a two-step Gibbs sampling procedure. The goal is to iteratively sample $\boldsymbol{\alpha}$ and the vector $(\rho, \beta, \sigma^2)$ based on the respective other full conditional distribution. As it is not straightforward to derive the conditional distribution for the latter vector, we will here derive conditional distributions for all parameters involved as if the parameter $\tau^2$ was fixed and later employ these distributions to derive a suitable proposal distribution in a Metropolis-Hastings procedure. In the following subsections we just state the (full) conditional distributions. A more detailed derivation can be found in \ref{appendix:FC_alpha}-\ref{appendix:FC_sigma2}.

\subsubsection{Full Conditional Distributions of $\boldsymbol{\alpha}$}

In the following we derive the full conditional distribution of $\boldsymbol{\widetilde{\alpha}}$.
It holds
\begin{align}
\label{eq:general_FC}
    p(\boldsymbol{\widetilde{\alpha}} | \beta, \sigma^2, \boldsymbol{x}) \propto p(\boldsymbol{x} |\boldsymbol{\widetilde{\alpha}}, \beta, \sigma^2) \cdot p(\boldsymbol{\widetilde{\alpha}}) = \mathcal{L}(\boldsymbol{\widetilde{\alpha}},\beta,\sigma^2)  \cdot p(\boldsymbol{\widetilde{\alpha}}).
\end{align}
Due to the conditional independence induced by the Markov assumption in the AR(1) model the likelihood of the parameters is given by
\begin{align}
\label{eq:Likelihood}
\mathcal{L}(\boldsymbol{\widetilde{\alpha}},\beta,\sigma^2) = p(\boldsymbol{x}|\alpha,\beta,\sigma^2) = \prod_{j=0}^{t-1} \phi(x_{t-j}|\alpha_{t-j} + \beta x_{t-j-1},\sigma^2), %\phi(x_0|0,Var(x_t)).
\end{align}
where $\phi(\cdot | \mu, \widetilde{\sigma}^2)$ denotes the density function of a normal distribution with expectation $\mu$ and variance $\widetilde{\sigma}^2$. Note, that we have assumed a degenerated distribution with point mass $1$ for the first entry in $\boldsymbol{x}$. An alternative option is to estimate the unconditional distribution. For increasing length of the time series the difference between these two approaches will however vanish. \\

With (\ref{eq:general_FC}) and (\ref{eq:Likelihood}) and the prior distributions specified in Section \ref{sec:modelSpec} the full conditional distributions of $\boldsymbol{\widetilde{\alpha}}$, can be derived analytically. Under the assumption that  $\boldsymbol{\widetilde{\theta}} = \boldsymbol{0}$ as specified in Section \ref{sec:LongRunMean} to regularize the long run mean, the full conditional distribution of $\boldsymbol{\widetilde{\alpha}}$ is given by
$$
\boldsymbol{\widetilde{\alpha}} | \beta,\sigma^2, \boldsymbol{x} \sim \mathcal{N}_t(\boldsymbol{\widetilde{\mu}_{post}},\boldsymbol{\tilde{\Sigma}_{post}}).
$$
with
\begin{equation*} 
\boldsymbol{\widetilde{\mu}_{post}} = \boldsymbol{\tilde{\Sigma}_{post}} \boldsymbol{\widetilde{\Delta}}\frac{1}{\sigma^2} \quad \text{and} \quad \boldsymbol{\tilde{\Sigma}_{post}} = \left(\boldsymbol{\tilde{\Sigma}}^{-1} + \frac{1}{\sigma^2}\begin{pmatrix}  \boldsymbol{I_t} & \boldsymbol{0} \\ \boldsymbol{0} & \boldsymbol{0} \end{pmatrix} \right)^{-1}.
\end{equation*} 
$\boldsymbol{\widetilde{\Delta}}$ in this case denotes
$$
\boldsymbol{\widetilde{\Delta}} = (x_2-\beta x_1, \dots, x_t - \beta x_{t-1}, 0, \dots, 0).
$$
As $\boldsymbol{\widetilde{\Delta}}$ incorporates data information up to time point (vector entry) $t$, is $0$ for time points $>t$ and $Cov(\alpha_{t+j}, \alpha_t) \longrightarrow 0$ with increasing $j$, the mean of the full conditional distribution tends to $0$, corresponding to the unconditional mean of the prior distribution. The covariance structure of the full conditional distribution behaves analogously. Therefore, the distribution of $\alpha_{t+j} \mid \boldsymbol{x}, \beta, \sigma^2$ in the long run tends to the prior distribution. This means that the prior distribution of $\boldsymbol{\alpha}$ effectively regularizes the distribution of $\boldsymbol{x}$ in the long horizon towards the prespecified long run mean and long run variance. 

Note that the derivations are independent of the specific choice of $\boldsymbol{\tilde{\Sigma}}$. If prior distribution assumptions for the parameters in $\boldsymbol{\tilde{\Sigma}}$ are used, we need to further condition on the hyper-parameters for the full conditional distribution of $\boldsymbol{\widetilde{\alpha}}$. 

\subsubsection{Full Conditional Distributions of $\rho, \beta, \sigma^2$}

If we assume an AR-covariance structure with prior distributions for its parameters as specified in Section \ref{sec:LongRunVariance}, the conditional distribution of $\rho$ is given by
 \begin{align*}
    \rho | \boldsymbol{\alpha}, \tau^2 \sim \mathcal{N}(\mu_{\rho,post}, \sigma^2_{\rho,post}, -1, 1)
\end{align*} 
where
\begin{align*}
    \sigma^2_{\rho,post} &= \left(\frac{\sum_{j=0}^{t-1}\alpha^2_{t-j-1}}{\tau^2} + \sigma^{-2}_{\rho} \right)^{-1} \\
    \mu_{\rho,post} &= \left(\frac{\sum_{j=0}^{t-1}\alpha_{t-j} \alpha_{t-j-1}}{\tau^2} + \frac{\mu_\rho}{\sigma^2_{\rho}}\right) \sigma^2_{\rho,post}.
\end{align*}
%As the prior distributions of $\tau^2$ is a degenerated distribution, the corresponding full conditional distribution is directly given.

The conditional distribution of $\beta$ is given by
\begin{align*}
    \beta | \boldsymbol{x}, \boldsymbol{\alpha}, \sigma^2 \sim \mathcal{N}\left(\mu_{\beta,post}, \sigma^2_{\beta,post}, -1, \sqrt{\frac{Var(x_t) - \sigma^2}{Var(x_t)}}\right)
\end{align*} 
where 
\begin{align*}
\sigma^2_{\beta,post} &= \left(\frac{\sum_{j=0}^{t-1}x^2_{t-j-1}}{\sigma^2} + (\sigma\sigma_{\beta})^{-2} \right)^{-1}\\
\mu_{\beta,post} &= \left(\frac{\sum_{j=0}^{t-1}\breve{d}_{t-j} x_{t-j-1}}{\sigma^2} + \frac{\mu_\beta}{\sigma^2\sigma^2_{\beta}}\right) \sigma^2_{\beta,post}.
\end{align*}
$\breve{d}_{t-j}$ is defined by $\breve{d}_{t-j} := x_{t-j}-\alpha_{t-j}$.
\\

The conditional distribution of $\sigma^2$ is given by an inverse gamma distribution with parameters 
$$
\tilde{a} = \frac{t+1}{2} + a \qquad \text{ and } \qquad \tilde{b} = \frac{\sum_{j=0}^{t-1} \epsilon^2_{t-j}}{2} + b + \frac{(\beta - \mu_{\beta})^2}{2\sigma_{\beta}}.
$$ 
This means
$$
\sigma^2|\boldsymbol{\widetilde{\alpha}},\beta, \boldsymbol{x} \sim \mathcal{IG}(\tilde{a},\tilde{b}).
$$
Note that this only holds if the prior of $\beta\mid\sigma^2$ is a normal distribution instead of a truncated normal distribution as assumed in Section \ref{sec:modelSpec}. When using a truncated distribution assumption, the derivation of the full conditional of $\sigma^2$ is more intricate as the prior distribution of $\beta$ also conditions on $\sigma^2$. Since our approach will make use of the full conditionals as proposal distributions in the Metropolis-Hastings part of our sampling routine, this simplification allows a more straightforward implementation while we observe that values outside the given truncation are highly unlikely and practically occur with zero probability in our application.

\subsection{Markov Chain Monte Carlo Inference} \label{sec:GibbsSampler}
In the following we assume again an AR-covariance structure for $\boldsymbol{\Sigma}$ determined by the parameters $\rho$ and $\tau^2$ with prior distributions as specified in Section \ref{sec:LongRunVariance}. To conduct inference, we use the Metropolis-Hastings within Gibbs sampler.
More specifically, we generate samples from the posterior distribution by iteratively sampling from the full conditional distribution of $\boldsymbol{\widetilde{\alpha}}$ given a sample of $(\rho, \beta, \sigma^2)$ and vice versa. Based on the derivation of the full conditional distribution for $\boldsymbol{\alpha}$ in the previous section we are able to directly sample from a multivariate normal distribution to generate values for $\boldsymbol{\alpha}$. To obtain a sample from $p(\rho, \beta, \sigma^2 \mid \boldsymbol{\alpha}, \boldsymbol{x})$ conditional on $\boldsymbol{\alpha}$, we apply the Metropolis-Hastings algorithm as neither the joint distribution of $\rho, \beta, \sigma^2$ nor each single full conditional distribution is available. A suitable and already available proposal distribution $q$ for these parameters is given by 
\begin{align}
\label{eq:proposal}
q(\rho, \beta, \sigma^2 \mid \boldsymbol{\alpha}, x) = q(\rho \mid \boldsymbol{\alpha}, x)q(\beta \mid \boldsymbol{\alpha}, \sigma^2, x) q(\sigma^2 \mid \boldsymbol{\alpha}, \beta, x).
\end{align}
In other words, we use the product of all full conditional distributions under the assumption of a fixed $\tau^2$. 

In the BTVC-AR(1) model we use this approach in a first step to draw from the joint posterior distribution $p(\boldsymbol{\widetilde{\alpha}}, \beta, \sigma^2 \mid \boldsymbol{x})$. A detailed description of the sampling routine can be found in \ref{appendix:algorithm}. In a second and final step, we use these samples to generate paths of the $x$-process as follows:
$$
x_{t+j}^{(m)} = \alpha_{t+j}^{(m)} + \beta^{(m)} x_{t+j-1}^{(m)}+ \epsilon_{t+j}, \qquad j > 0.
$$

%%%% PART TWO 
 
\section{Application To Interest Rate Data}
\label{sec:Application}
We now apply the BTVC-AR(1) model to the first principal component (PC) of a principal component analysis (PCA) on interest rate data to predict the term structure of interest rates and compare it to the \textit{2-Additive-Factor Gaussian (Gauss2++) model} (see, e.g., \citet{Brigo2007}) and the \textit{dynamic Nelson-Siegel model} (see \citet{diebold2006}) with respect to the forecasting performance and the long run distribution. 

\subsection{Motivation and Background}

The Gauss2++ model is a popular short-rate model in the insurance industry, used, e.g., to classify certified pension contracts into risk classes. Because its mean reversion level is calibrated to external interest rate forecasts, it generates realistic interest rates in the long horizon, which is a necessary model feature for insurance companies, as they are obliged to calculate risk measures and performance scenarios for specific insurance contracts for up to 40 years (see \citet{european2017}). Nevertheless, \citet{diebold2006} point out that short-rate models perform poorly in forecasting. Their dynamic Nelson-Siegel model shows a better forecasting performance than the Gauss2++ model in the short horizon, but can produce unrealistic interest rates in the very long horizon. Our model, which we call the \textit{BTVC-AR(1)-Factor model} in the following as it applies the BTVC-AR(1) model to the first PC of a PCA, combines both: a good forecasting performance in the short horizon and realistic interest rates in the long horizon. It further accounts for the strong autocorrelation and the (close to) random walk behavior of interest rates.

\subsection{Data}
\begin{center}
\begin{figure}
\includegraphics[scale=0.7]{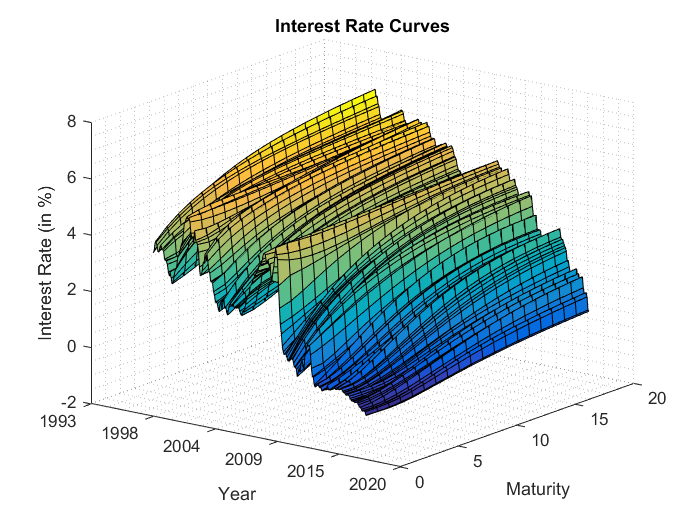}
\caption{Time series of the term structure of German government bond yields.}
\label{TSOfIRC}
\end{figure}
\end{center}

\begin{center}
\begin{figure}
\includegraphics[scale=0.5]{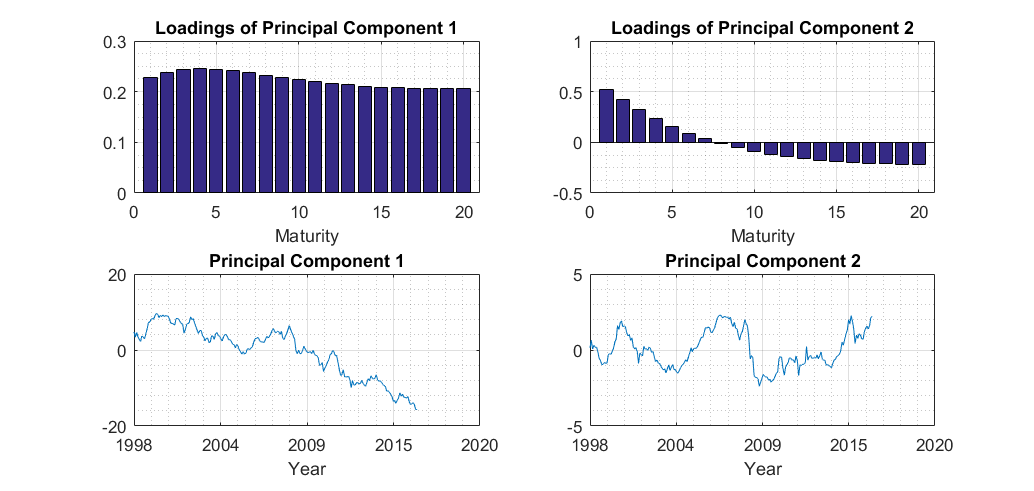}
\caption{The scores and the loadings of the first two PCs.}
\label{fig:LoadingsScores}
\end{figure}
\end{center}

We use data of the German term structure of interest rates estimated by the Deutsche Bundesbank from prices of German government bonds. The exact estimation procedure can be found in \citet{schich1997}. The time span ranges from September 1997 to August 2016. Figure \ref{TSOfIRC} shows the monthly evolution of the interest rate curves.
In the last ten to fifteen years a decrease of the interest rates can be observed. Each maturity represents a dimension in the data set. We use PCA to reduce the dimension of the data set for the following reason. According to \citet{litterman1991} a three factor model can explain for each interest rate with a specific maturity a minimum of $96\%$ of the variability in the data. We here extract these (principle) factors but only use the first two to facilitate a fair comparison with the Gauss2++ model, which is a two factor model. Furthermore, the first two PCs already account for more than $99\%$ of the variability in the given data. Figure \ref{fig:LoadingsScores} shows the loadings and the time series of the two extracted PCs.
The loadings of the first PC are similar for all 20 maturities, while the loadings of the second PC are positive for short and negative for long maturities. The first and the second PC are therefore often interpreted as level and slope of the term structure, respectively.

The decrease of the interest rates in the last years is also visible in the level factor, showing a downward trend. There is an ongoing discussion in the literature about mean reversion of interest rates. Economic theory predominantly assumes that interest rates are (in the long run) mean reverting. But statistical evidence is not so clear (\citet{vanDenEnd2011}). The mainstream literature says that unit roots can not be rejected, which would imply that interest rates are not mean reverting (see, e.g, \citet{Campbell1991, Rose1988, siklos1997, Stock1988}). More recent literature investigates the unit root hypothesis by fractional integrated techniques that apply differencing to time series by an order smaller than or greater than one (see, e.g., \citet{Bekdache2000, Gil2004}). These studies find that shocks to interest rates have a long memory, which explains their (close to) random walk behavior.

\subsection{Estimation of Model Parameters}
\label{sec:ParameterEstimation}
In this subsection the estimation of the BTVC-AR(1)-Factor model and the two benchmark models is described.

\subsubsection{Modeling Interest Rates with the BTVC-AR(1)-Factor Model}
The factors of our BTVC-AR(1)-Factor model are the first two PCs extracted by a PCA and interpreted as level and slope of the interest rate curve. The level factor shows a (close to) random walk behavior, which can not be adequately captured by a stationary linear model. Following the economic theory view that interest rates (and therefore also the level) are mean reverting (in the long run) and assuming that the random walk behavior results from changes in the mean reversion level, we use therefore the BTVC-AR(1) model for this PC. It allows us to account for the (close to) random walk behavior as well as to regularize the level of the interest rate curve in the long horizon via prior assumptions. The slope factor is more stable over time. As an augmented Dickey Fuller test suggests that the existence of a unit root can be rejected, a linear AR(1) model is used for this factor. By modeling the level and the slope factor interest rate forecasts $\hat{r}_t(\tau)$ with maturity $\tau$ can be calculated via
\begin{align}
\hat{r}_t(\tau) &= \mu(\tau) + \xi_1(\tau) \hat{l}_t + \xi_2(\tau) \hat{s}_t,
\end{align}
where $\hat{l}_t$ and $\hat{s}_t$ denote the forecasts of the level and the slope factor, respectively. $\xi_1(\tau)$ and $\xi_2(\tau)$ denote the loading of the first and second PC for maturity $\tau$. Before applying the PCA the data has been centered and therefore $\mu(\tau)$ is the mean interest rate of the data set for maturity $\tau$. We now specify the prior assumptions of the BTVC-AR(1) model for the level factor and the estimation procedure of the AR(1) model for the slope factor. 
\\
\\
\\
\textit{The Level Factor}\vspace{0.4cm}\\
\noindent \onehalfspacing
\textbf{Latent AR1 constant }$\boldsymbol{\alpha}$. For this application we assume an AR-covariance structure for the $\alpha$-process of the BTVC-AR(1) model with the parameters $\rho$ and $\tau^2$ representing the correlation of two successive time points and the conditional variance, respectively. The unconditional mean of the $\alpha$-process is set to $0$, which implies the assumption that the long run mean of the level factor is $0$. Because we also assume that the slope factor is a centered process this means that the long run interest rate curve converges in expectation to the average interest rate curve of the dataset. 
\\
\textbf{Autocorrelation parameter }$\boldsymbol{\rho}$. As specified in Section \ref{sec:LongRunVariance} we assume for $\rho$ a truncated normal distribution with the parameters $\mu_{\rho} = 0.98$ and $\sigma^2_{\rho} = 0.001^2$ with lower truncation $-1$ and and upper truncation $1$ as a hyper prior, i.e.,
\begin{align*}
    \rho \sim \mathcal{N}(0.98,0.001,-1,1)
\end{align*}
The truncation ensures the stationarity of the process. The parameters of this hyper-prior rely on expert judgment and incorporate the assumption of a weak mean reverting $\alpha$-process into the model and therefore allow the mean reversion level of the level factor to deviate from the long run mean for longer periods. This yields the (close to) random walk behavior present in (our) interest rate data. 
\\
\textbf{Variance of the latent process}. According to Section \ref{sec:LongRunVariance} the parameter $\tau^2$ is set in each iteration of the sampling procedure such that the long run variance of the level factor amounts to a prespecified value. We here use the value $120$, which is inferred from a quantile of the unconditional distribution. By giving consideration of the rather unusual market situation of extremely low interest rates we make the assumption that the last observation is equal to the $7.5\%$-quantile. Due to the model assumptions, the unconditional distribution is normal with mean $0$ and the corresponding unconditional variance can be calculated easily. 
\\
\textbf{Slope parameter of the AR(1) model}. For $\beta$ we assume that $\mu_{\beta} = 0.95$ and $\sigma^2_{\beta} = 0.015^2$. This expert judgment represents a weak mean reversion to the time-varying mean reversion level. The lower and upper truncation of the truncated normal distribution amount to $-1$ and $\sqrt{\frac{Var(x_t) - \sigma^2}{Var(x_t)}}$ to ensure the stationarity of the model as well as the positivity of $\tau^2$, i.e.,
\begin{align*}
    \beta|\sigma^2 \sim \mathcal{N}\left(0.95,\sigma^2 0.015^2, -1,\sqrt{\frac{Var(x_t) - \sigma^2}{Var(x_t)}}\right)
\end{align*}
\textbf{Residual variance}. For the prior distribution of $\sigma^2$ the shape and scale parameter $a$ and $b$ are set to $0.5$ and $2$ respectively, representing an uninformative prior.
\\

By specifying the parameters of the prior (and hyper-prior) distributions the full conditional distribution of $\boldsymbol{\widetilde{\alpha}}$ as well as the conditional distributions of the other parameters can be analytically derived as described in Section \ref{sec:BayesianInference}. Combining the Gibbs Sampler and the Metropolis-Hastings algorithm as explained in Section~\ref{sec:GibbsSampler}, paths of the level factor can be generated. Forecasts of the level factor are then represented by the average of the simulated paths.
\\
\\
\\
%\subsubsection*{The slope factor}
\textit{The Slope Factor} \vspace{0.4cm} \\
\onehalfspacing \noindent The linear AR(1) model for the slope factor is given by \begin{align*}
    s_t = c + \gamma s_{t-1} + \eta_t,    
\end{align*}
where $\gamma$ is a real valued constant between $-1$ and $1$ and $\eta_t$ is a Gaussian white noise process, i.e., $\eta_t \sim \mathcal{N}(0,\breve{\sigma}^2)$. The constant parameter $c$ is set to $0$.
The other parameters are estimated by a standard ordinary least squares approach.

\subsubsection{Modeling Interest Rates With the Gauss2++ Model}
The Gauss2++ model -- in a different representation also known as the 2-Factor-Hull-White model -- is a popular interest rate model in the insurance industry used for pricing interest rate derivatives as well as for risk management and forecasting purposes. The model assumes that the short-rate $r(t)$, which is the interest rate with an infinitesimal small maturity, is given by the sum of two latent processes $(x(t))_{t\geq 0}$ and $(y(t))_{t\geq 0}$, and a deterministic function $\varphi$:
$$
r(t) = x(t) + y(t) + \varphi(t).
$$
The latent processes are modeled by dependent Ornstein-Uhlenbeck processes, which are the continuous version of a linear AR(1) process. Interest rates with longer maturities are then derived from the short-rate via pricing the corresponding zero-coupon bonds, which is analytically possible due to the model's distributional assumptions.\\
The estimation process is materially different from the one of the other two models as it does not use historic data but calibrates the model to current future market assumptions (implicitly) provided by the current interest rate curve, interest rate derivatives as well as interest rate forecasts. By applying the downhill simplex algorithm the parameters of the model are chosen in such a way that forward rates -- implicitly given by the current interest rate curve -- and swaption prices are met in expectation. The relevant data has been extracted from Bloomberg. Additionally the mean reversion level of the two latent factors are analytically set such that two interest rate forecasts with a maturity of 3 months and 10 years, which are published by the OECD, are met in expectation. This approach is in line with the standard calibration procedure in the insurance industry (see, e.g., \citet{KornWagner2020}).

\subsubsection{Modeling Interest Rates With the Dynamic Nelson-Siegel Model}
The dynamic Nelson-Siegel model of \citeauthor{diebold2006} applies specific time series models to extracted latent factors. Diebold and Li tested several time series models on the level, slope and curvature factors of the Nelson-Siegel interest rate curve and compared the forecasting performance [Diebold and Li, 2006]. In this paper we follow one of their approaches, in which they apply a PCA on interest rate data and use an univariate linear AR(1) process for each of the first three PCs. Because of comparison reasons to the other two two-factor models in this paper, we just use the first two PCs. The parameters of the AR(1) model are estimated by the ordinary least squares method.

\subsection{Backtest}
We now compare the forecasting performance of the BTVC-AR(1)-Factor model, the Gauss2++ model and the dynamic Nelson-Siegel model and analyse their long run distributions of the 10-year interest rate.

\subsubsection{Comparison of the Forecasting Performance}
For the out-of-sample backtest we apply an expanding window approach. The data of the first 10 years of the observations are used to estimate the parameters of the BTVC-AR(1)-Factor model and the dynamic Nelson-Siegel model as described in the Section \ref{sec:ParameterEstimation}. The Gauss2++ model is calibrated to the current market data. We then forecast the interest rates for the maturities of 1, 3, 5 and 10 years (representing the yield curve) for the horizons of 1, 3, 6 and 12 months. We expand the training sample by one month and repeat the procedure again. This is done until 12 months before the last observation in the data set. To evaluate the forecasting performance the error between the predicted interest rate $\hat{r}_{\tau}(t)$ and the actual interest rate $r_{\tau}(t)$ with the maturity $\tau$ is calculated, i.e.,
\begin{align*}
    error_{\tau}(t) = r_{\tau}(t) - \hat{r}_{\tau}(t).
\end{align*}
Table (\ref{tbl:1monthFC})-(\ref{tbl:12monthFC}) in the \ref{appendix:Tables} show the mean and the standard deviation of this error for each model. In addition, the root mean squared error
\begin{align}
RMSE(\tau) = \sqrt{\frac{1}{N}\sum_{k=1}^N (r_{\tau}(k)-\hat{r}_{\tau}(k))^2}
\end{align}
for the given deviation is calculated, where $N$ is the number of forecasts conducted in the backtest.
\\

The RMSE for the 1-month ahead forecasts is similar for all three models. For longer forecasting horizons the Gauss2++ model shows the highest RMSE. For example, the 6-month ahead forecast of the 10-year interest rate of the Gauss2++ model has a RMSE, which is approximately twice as high as the RMSE of the other two models and more than three times as high for the 12-month ahead forecast. This supports the statement of \citet{diebold2006} that short-rate models perform poorly in forecasting. However, it should be mentioned that the performance of the Gauss2++ model highly depends on the interest rate forecasts used in the calibration process. Regarding the predominant negative mean error suggests that the OECD forecasts have been too optimistic in the past.

The results of the BTVC-AR(1)-Factor model and the dynamic Nelson-Siegel model are more consistent. For the forecasting horizon of 1-month the BTVC-AR(1)-Factor model shows a slightly lower RMSE except for the 10-year interest rate. For the 3-months, 6-months and 12-months forecasting horizons the BTVC-AR(1) model shows a lower RMSE for the short maturities, but a higher RMSE for the longer maturities compared to the dynamic Nelson-Siegel model. Note that the dynamic Nelson-Siegel model anticipated the downward trend present in the last years, which might have been beneficial in terms of the forecasting performance in the past, but also produces unrealistic interest rates in the long horizon. In contrast the BTVC-AR(1)-Factor model forces the model to mean revert to a prespecified level to regularize the interest rates in the long horizon. It can therefore follow the current trend only for a couple of time steps, which might explain the slightly worse performance for the 6- and 12-months forecasting horizon. The fact that the RMSE error is still similar to the dynamic Nelson-Siegel model suggests that this does not affect the forecasting performance in the short horizon much.

\subsubsection{Comparison of the Distribution in the Long Run}
We further investigate the interest rate distribution in the long horizon. This is especially important for insurance companies as risk measures and performance scenarios for their products have to be calculated for up to 40 years (see, e.g., \citet{european2017}). We therefore fit all three models on all data points up to the last observation date of the data set. We then simulate paths of the 10-year interest rate and visualize the distribution in 40 years. 
\begin{figure}
\begin{center}
\includegraphics[scale=0.6]{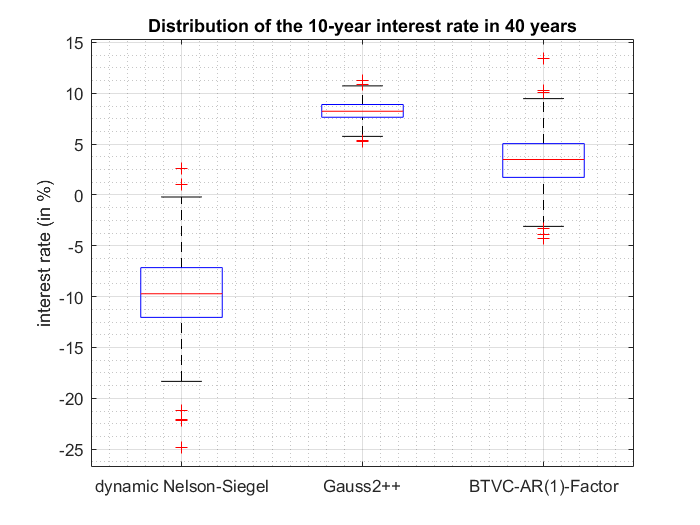}
\caption{Comparison of the distributions of the 10-year interest rate in 40 years modeled by the dynamic Nelson-Siegel model, the Gauss2++ model and the BTVC-AR(1)-Factor model.}
\end{center}
\end{figure}
The median of the dynamic Nelson-Siegel model amounts to approximately -10\%. A value that is not realistic for the 10-year interest rate. In comparison, the distribution of the BTVC-AR(1)-Factor model and the Gauss2++ model seem to be more realistic as the range of their distributions is (mainly) positive between $0\%$ and $10\%$. It can be observed that the standard deviation of the Gauss2++ model is much smaller than of the BTVC-AR(1)-Factor model and as the median is quite high negative values are not reached by this model. This is due to the fact that the Gauss2++ model assumes a stronger mean reversion than historic data would suggest. The (close to) random walk behavior is better captured by the BTVC-AR(1)-Factor model leading to a prediction range which fits historical observations quite well. This is due to the regularization of the mean and the standard deviation of the BTVC-AR(1)-Factor model induced by appropriate prior assumptions, which represents the main difference to other interest rate models.

\section{Conclusion} \label{sec:Conclusion}

In this paper we introduced a new Bayesian framework for the TVC-AR(1) model particularly suitable for nearly integrated time series which can not be estimated by a linear model consistent with economic theory or historical observations. In these cases a (close to) random walk behavior can be an indication for a missing variable, for which we account for by the usage of a non-linear model. The time-varying constant of the BTVC-AR(1) allows a stochastic mean reversion level leading to realizations, which exhibit a random walk behavior although being stationary and do not have an exploding long run variance. Additionally, with the Bayesian approach it is possible to incorporate prior assumption about the long run distribution into the model without affecting the short-term predictions adversely. This gives the possibility to include expert knowledge or well known economic facts about the long-term behavior of the time series into the model that is otherwise fully data-driven in the short term forecast.

We apply the proposed approach to interest rate data. We find that the BTVC-AR(1)-Factor model, which applies a BTVC-AR(1) model to the first PC of a PCA, shows a similar forecasting performance as the dynamic Nelson-Siegel model in the short horizon but in contrast produces realistic interest rates in the very long horizon and also yields better forecasts compared to the Gauss2++ model.

The presented framework allows for many different specifications and is, in particular, flexible in terms of the assumed covariance structure of the latent $\alpha$ process in the model. In this paper we propose an AR-covariance structure and explain how model parameters can be inferred in this special case. Investigating other covariance structures may further improve the forecasting performance in the short horizon while still regularizing the distribution in the long run.

\newpage
\section*{References}
%\nocite{*}
\bibliographystyle{plainnat}
\bibliography{Literatur}

\begin{thebibliography}{37}
\providecommand{\natexlab}[1]{#1}
\providecommand{\url}[1]{\texttt{#1}}
\expandafter\ifx\csname urlstyle\endcsname\relax
  \providecommand{\doi}[1]{doi: #1}\else
  \providecommand{\doi}{doi: \begingroup \urlstyle{rm}\Url}\fi

\bibitem[Bacon and Watts(1971)]{Bacon1971}
David~W. Bacon and Donald~G. Watts.
\newblock Estimating the transition between two intersecting straight lines.
\newblock \emph{Biometrika}, 58\penalty0 (3):\penalty0 525--534, 1971.

\bibitem[Baum et~al.(2000)Baum, Bekdache, et~al.]{Bekdache2000}
Christopher~F. Baum, Basma Bekdache, et~al.
\newblock A re-evaluation of empirical tests of the fisher hypothesis.
\newblock \emph{Working Papers in Economics}, page 148, 2000.

\bibitem[Bekaert et~al.(2010)Bekaert, Cho, and Moreno]{bekaert2010}
Geert Bekaert, Seonghoon Cho, and Antonio Moreno.
\newblock New keynesian macroeconomics and the term structure.
\newblock \emph{Journal of Money, Credit and Banking}, 42\penalty0
  (1):\penalty0 33--62, 2010.

\bibitem[Brigo and Mercurio(2007)]{Brigo2007}
Damiano Brigo and Fabio Mercurio.
\newblock \emph{Interest rate models -- theory and practice: with smile,
  inflation and credit}.
\newblock Springer Science \& Business Media, 2007.

\bibitem[Campbell and Shiller(1991)]{Campbell1991}
John~Y. Campbell and Robert~J. Shiller.
\newblock Yield spreads and interest rate movements: A bird's eye view.
\newblock \emph{The Review of Economic Studies}, 58\penalty0 (3):\penalty0
  495--514, 1991.

\bibitem[Chan and Tong(1986)]{Chan1986}
Kung~Sik Chan and Howell Tong.
\newblock On estimating thresholds in autoregressive models.
\newblock \emph{Journal of Time Series Analysis}, 7\penalty0 (3):\penalty0
  179--190, 1986.

\bibitem[Chib(2001)]{Chib2001}
Siddhartha Chib.
\newblock Markov chain monte carlo methods: computation and inference.
\newblock In \emph{Handbook of Econometrics}, volume~5, pages 3569--3649.
  Elsevier, 2001.

\bibitem[Cieslak and Povala(2015)]{cieslak2015}
Anna Cieslak and Pavol Povala.
\newblock Expected returns in treasury bonds.
\newblock \emph{The Review of Financial Studies}, 28\penalty0 (10):\penalty0
  2859--2901, 2015.

\bibitem[Dickey and Fuller(1979)]{DickeyFuller1979}
David~A Dickey and Wayne~A Fuller.
\newblock Distribution of the estimators for autoregressive time series with a
  unit root.
\newblock \emph{Journal of the American statistical association}, 74\penalty0
  (366a):\penalty0 427--431, 1979.

\bibitem[Diebold and Li(2006)]{diebold2006}
Francis~X. Diebold and Canlin Li.
\newblock Forecasting the term structure of government bond yields.
\newblock \emph{Journal of Econometrics}, 130\penalty0 (2):\penalty0 337--364,
  2006.

\bibitem[Duffee(2011)]{Duffee2011}
Gregory~R. Duffee.
\newblock Forecasting with the term structure: The role of no-arbitrage
  restrictions.
\newblock Technical report, Working Paper, 2011.

\bibitem[{European Union}(2017)]{european2017}
{European Union}.
\newblock Commission delegated regulation (eu) 2017/653 of 8 march 2017
  supplementing regulation (eu) no 1286/2014 of the european parliament and of
  the council on key information documents for packaged retail and
  insurance-based investment products (priips) by laying down regulatory
  technical standards with regard to the presentation, content, review and
  revision of key information documents and the conditions for fulfilling the
  requirement to provide such documents.
\newblock \emph{Official Journal of the European Union}, 60, 2017.

\bibitem[Gelman et~al.(2013)Gelman, Carlin, Stern, Dunson, Vehtari, and
  Rubin]{gelman2013bayesian}
Andrew Gelman, John~B. Carlin, Hal~S. Stern, David~B. Dunson, Aki Vehtari, and
  Donald~B. Rubin.
\newblock \emph{Bayesian data analysis}.
\newblock Chapman and Hall/CRC, 2013.

\bibitem[Gil-Alana(2004)]{Gil2004}
Luis~A. Gil-Alana.
\newblock Long memory in the us interest rate.
\newblock \emph{International Review of Financial Analysis}, 13\penalty0
  (3):\penalty0 265--276, 2004.

\bibitem[Goldfeld and Quandt(1973)]{Goldfeld1973}
Stephen Goldfeld and Richard Quandt.
\newblock The estimation of structural shifts by switching regressions.
\newblock In \emph{Annals of Economic and Social Measurement, Volume 2, number
  4}, pages 475--485. NBER, 1973.

\bibitem[Granger and Andersen(1978)]{Granger1978}
C.~W.~J. Granger and A.~P. Andersen.
\newblock \emph{An introduction to bilinear time series models}.
\newblock Vandenhoeck and Ruprecht: Göttingen, 1978.

\bibitem[Haggan and Ozaki(1981)]{Haggan1981}
Val{\'e}rie Haggan and Tohru Ozaki.
\newblock Modelling nonlinear random vibrations using an amplitude-dependent
  autoregressive time series model.
\newblock \emph{Biometrika}, 68\penalty0 (1):\penalty0 189--196, 1981.

\bibitem[Hamilton(1989)]{Hamilton1989}
James~D. Hamilton.
\newblock A new approach to the economic analysis of nonstationary time series
  and the business cycle.
\newblock \emph{Econometrica: Journal of the Econometric Society}, pages
  357--384, 1989.

\bibitem[Jansen and Ter{\"a}svirta(1996)]{Jansen1996}
Eilev~S. Jansen and Timo Ter{\"a}svirta.
\newblock Testing parameter constancy and super exogeneity in econometric
  equations.
\newblock \emph{Oxford Bulletin of Economics and Statistics}, 58\penalty0
  (4):\penalty0 735--763, 1996.

\bibitem[Korn and Wagner(2019)]{KornWagner2020}
Ralf Korn and Andreas Wagner.
\newblock \emph{Praxishandbuch Lebensversicherungsmathematik : Simulation und
  Klassifikation von Produktent}.
\newblock VVW GmbH, 2019.

\bibitem[Kozicki and Tinsley(2001)]{Kozicki2001}
Sharon Kozicki and Peter~A. Tinsley.
\newblock Shifting endpoints in the term structure of interest rates.
\newblock \emph{Journal of Monetary Economics}, 47\penalty0 (3):\penalty0
  613--652, 2001.

\bibitem[Lanne and Saikkonen(2002)]{Lanne2002}
Markku Lanne and Pentti Saikkonen.
\newblock Threshold autoregressions for strongly autocorrelated time series.
\newblock \emph{Journal of Business \& Economic Statistics}, 20\penalty0
  (2):\penalty0 282--289, 2002.

\bibitem[Lim and Tong(1980)]{Tong1980}
K.~S. Lim and H.~Tong.
\newblock Threshold autoregressions, limit cycles, and data.
\newblock \emph{Journal of the Royal Statistical Sociaty, B}, 42:\penalty0
  245--92, 1980.

\bibitem[Litterman and Scheinkman(1991)]{litterman1991}
Robert Litterman and Jose Scheinkman.
\newblock Common factors affecting bond returns.
\newblock \emph{Journal of Fixed Income}, 1\penalty0 (1):\penalty0 54--61,
  1991.

\bibitem[Millar and Meyer(2000)]{Millar2000}
Russell~B. Millar and Renate Meyer.
\newblock Non-linear state space modelling of fisheries biomass dynamics by
  using metropolis-hastings within-gibbs sampling.
\newblock \emph{Journal of the Royal Statistical Society: Series C (Applied
  Statistics)}, 49\penalty0 (3):\penalty0 327--342, 2000.

\bibitem[Morley(2009)]{Morley2009}
James~C. Morley.
\newblock Nonlinear time series in macroeconomics.
\newblock \emph{Encyclopedia of Complexity and System Science, forthcoming},
  2009.

\bibitem[Quandt(1958)]{Quandt1958}
Richard~E. Quandt.
\newblock The estimation of the parameters of a linear regression system
  obeying two separate regimes.
\newblock \emph{Journal of the American Statistical Association}, 53\penalty0
  (284):\penalty0 873--880, 1958.

\bibitem[Rao and Gabr(2012)]{Rao1984}
T.~Subba Rao and M.~M. Gabr.
\newblock \emph{An introduction to bispectral analysis and bilinear time series
  models}, volume~24.
\newblock Springer Science \& Business Media, 2012.

\bibitem[Rose(1988)]{Rose1988}
Andrew~K. Rose.
\newblock Is the real interest rate stable?
\newblock \emph{The Journal of Finance}, 43\penalty0 (5):\penalty0 1095--1112,
  1988.

\bibitem[Rudebusch and Wu(2008)]{rudebusch2008}
Glenn~D. Rudebusch and Tao Wu.
\newblock A macro-finance model of the term structure, monetary policy and the
  economy.
\newblock \emph{The Economic Journal}, 118\penalty0 (530):\penalty0 906--926,
  2008.

\bibitem[Schich(1997)]{schich1997}
Sebastian Schich.
\newblock Sch{\"a}tzung der deutschen zinsstrukturkurve.
\newblock \emph{Bundesbank Series 1 Discussion Paper}, 1997.

\bibitem[Siklos and Wohar(1997)]{siklos1997}
Pierre~L. Siklos and Mark~E. Wohar.
\newblock Convergence in interest rates and inflation rates across countries
  and over time.
\newblock \emph{Review of International Economics}, 5\penalty0 (1):\penalty0
  129--141, 1997.

\bibitem[Stock and Watson(1988)]{Stock1988}
James~H. Stock and Mark~W. Watson.
\newblock Testing for common trends.
\newblock \emph{Journal of the American Statistical Association}, 83\penalty0
  (404):\penalty0 1097--1107, 1988.

\bibitem[Ter{\"a}svirta(1994)]{Terasvirta1994}
Timo Ter{\"a}svirta.
\newblock Specification, estimation, and evaluation of smooth transition
  autoregressive models.
\newblock \emph{Journal of the American Statistical Association}, 89\penalty0
  (425):\penalty0 208--218, 1994.

\bibitem[Ter{\"a}svirta et~al.(2010)Ter{\"a}svirta, Tj{\o}stheim, Granger,
  et~al.]{Terasvirta2010}
Timo Ter{\"a}svirta, Dag Tj{\o}stheim, C.~W.~J. Granger, et~al.
\newblock \emph{Modelling nonlinear economic time series}.
\newblock Oxford University Press Oxford, 2010.

\bibitem[van~den End(2011)]{vanDenEnd2011}
Jan~Willem van~den End.
\newblock Statistical evidence on the mean reversion of interest rates.
\newblock \emph{De Nederlandsche Bank Working Paper}, 2011.

\bibitem[Van~Dijk et~al.(2014)Van~Dijk, Koopman, Van~der Wel, and
  Wright]{Dijk2014}
Dick Van~Dijk, Siem~Jan Koopman, Michel Van~der Wel, and Jonathan~H. Wright.
\newblock Forecasting interest rates with shifting endpoints.
\newblock \emph{Journal of Applied Econometrics}, 29\penalty0 (5):\penalty0
  693--712, 2014.

\end{thebibliography}

\newpage
\appendix
\section{Full Conditional Distributions}

\subsection{The Full Conditional Distribution of $\boldsymbol{\widetilde{\alpha}}$}
\label{appendix:FC_alpha}
The prior distribution of $\boldsymbol{\widetilde{\alpha}}$ is a centered Gaussian process with a specific covariance structure $\boldsymbol{\widetilde{\Sigma}}$, i.e.,
\begin{align*}
\boldsymbol{\widetilde{\alpha}} = (\alpha_1, ... , \alpha_t, ..., \alpha_{t+h}) \sim \mathcal{N}_t(\boldsymbol{0}, \boldsymbol{\widetilde{\Sigma}})
\end{align*}
The following derivations will be independent of the specific choice of $\boldsymbol{\widetilde{\Sigma}}$. By defining 
%$Var(\alpha_s) = \tau^2$  $\forall s\geq 1$. By defining 
$$
\Delta_j = x_{j+1} - \beta x_{j}
$$ 
as well as $\boldsymbol{\Delta} = (\Delta_{0}, \ldots, \Delta_{t-1})^\top$ and the fact that 
$$
\phi(x_{t}|\alpha_{t} + \beta x_{t-1},\sigma^2) = \phi(\alpha_{t}|\Delta_{t-1},\sigma^2) 
$$allows a straightforward derivation of the full conditional of $\boldsymbol{\widetilde{\alpha}}$: 
%is equivalent to the likelihood of $\alpha \sim \mathcal{N}_t(\Delta, \sigma^2 I)$. 
\begin{align*}
    p(\boldsymbol{\widetilde{\alpha}} | \beta, \sigma^2, \boldsymbol{x}) &\propto p( \boldsymbol{x} | \boldsymbol{\widetilde{\alpha}}, \beta, \sigma^2) p(\boldsymbol{\widetilde{\alpha}}|\beta, \sigma^2) \\
    &\propto p( \boldsymbol{x} | \boldsymbol{\alpha}, \beta, \sigma^2) p(\boldsymbol{\widetilde{\alpha}}) \\
    &\propto \exp\left(-\frac{1}{2\sigma^2} (\boldsymbol{\alpha} - \boldsymbol{\Delta})^\top (\boldsymbol{\alpha} - \boldsymbol{\Delta})\right) \cdot \exp\left(-\frac{1}{2} \boldsymbol{\widetilde{\alpha}}^\top \boldsymbol{\widetilde{\Sigma}^{-1}} \boldsymbol{\widetilde{\alpha}}\right)\\
    &\propto \exp\left( -\frac{1}{2} (  \boldsymbol{\widetilde{\alpha}}^\top \boldsymbol{\widetilde{\Sigma}^{-1}_{post}} \boldsymbol{\widetilde{\alpha}} - 2 \boldsymbol{\widetilde{\alpha}^\top} \boldsymbol{\widetilde{\Sigma}^{-1}_{post}} \underset{=:\boldsymbol{\widetilde{\mu}_{post}}}{\underbrace{\boldsymbol{\widetilde{\Sigma}_{post}} \boldsymbol{\widetilde{\Delta}_0} \frac{1}{\sigma^2}} } ) \right) 
\end{align*}
with $\boldsymbol{\widetilde{\Sigma}^{-1}_{post}} = \boldsymbol{\widetilde{\Sigma}^{-1}} + \frac{1}{\sigma^2} \begin{pmatrix} \boldsymbol{I_t} & \boldsymbol{0} \\ \boldsymbol{0} & \boldsymbol{0} \end{pmatrix}$ and $\boldsymbol{\widetilde{\Delta}_0} = (\boldsymbol{\Delta^\top}, \boldsymbol{0})^\top$. 
\\

This is the kernel of a multivariate Gaussian distribution with covariance $\boldsymbol{\widetilde{\Sigma}_{post}}$ and mean vector $\boldsymbol{\widetilde{\mu}_{post}}$, i.e
 $$
\boldsymbol{\widetilde{\alpha}}\mid \beta,\sigma^2, \boldsymbol{x} \sim \mathcal{N}(\boldsymbol{\widetilde{\mu}_{post}}, \boldsymbol{\widetilde{\Sigma}_{post}}).
$$

\subsection{The Full Conditional Distribution of $\boldsymbol{\rho}$}
If an AR-covariance structure is assumed for $\boldsymbol{\widetilde{\alpha}}$ the latent $\alpha$-process can be written in the following form
$$
\alpha_t = \rho \alpha_{t-1} + \eta_t,
$$
where $\rho$ determines the correlation between two successive time steps and $\eta_t$ is a Gaussian white noise process, i.e., $\eta_t \sim \mathcal{N}(0,\tau^2)$.

The full conditional distribution of $\rho$ can be therefore derived as follows:
\begin{equation}
\label{eq:rho_FC}
p(\rho | \tau^2, \boldsymbol{\alpha}) \propto \mathcal{L}(\rho,\tau^2) \cdot p(\rho) = \prod_{j=0}^{t-1} \phi(\alpha_{t-j}|\rho \alpha_{t-j-1},\tau^2) \cdot p(\rho).
\end{equation}
The likelihood $\mathcal{L}(\cdot)$ in the above equation can be reformulated as
\begin{align*}
\mathcal{L} &\propto \exp\left(-\frac{1}{2\tau^2} \left\{-2\rho\left[\sum_{j=0}^{t-1}\alpha_{t-j} \alpha_{t-j-1}\right]+\rho^2 \left[\sum_{j=0}^{t-1}\alpha^2_{t-j-1}\right]\right\}\right).
\end{align*}
The calculation is similar to the one in appendix \ref{appendix:ReWriteLikelihood}. Defining the two terms in square brackets as $\eta$ and $\chi$, respectively, we get 
$$
\mathcal{L} \propto \exp\left(-\frac{1}{2\tau^2} \left\{-2\rho\eta+\rho^2 \chi\right\}\right).
$$
Plugging this into (\ref{eq:rho_FC}) and using a normal prior with parameters $\mu_{\rho}, \sigma^2_{\rho}$ for $\rho$, we have
\begin{align*}
p(\rho|\tau^2,\boldsymbol{\alpha}) &\propto \exp\left(-\frac{1}{2}\left\{ \frac{\rho^2 \chi}{\tau^2} - 2\frac{\rho\eta}{\tau^2} \right\}\right) \exp\left(-\frac{1}{2}\left\{ \frac{\rho^2}{\sigma^2_{\rho}} - 2\frac{\rho\mu_{\rho}}{\sigma^2_{\rho}} \right\}\right)\\
&\propto \exp\left(-\frac{1}{2}\left\{\rho^2 \cdot \left(\frac{\chi}{\tau^2} + \sigma^{-2}_{\rho} \right) -2\rho\left(\frac{\eta}{\tau^2} + \frac{\mu_\rho}{\sigma^2_{\rho}}\right) \right\}\right)
\end{align*}
and thus $\rho | \tau^2, \boldsymbol{\alpha} \sim \mathcal{N}(\mu_{\rho,post}, \sigma^2_{\rho,post})$ with
$$
\sigma^2_{\rho,post} = \left(\frac{\chi}{\tau^2} + \sigma^{-2}_{\rho} \right)^{-1}$$ and $$\mu_{\rho,post} = \left(\frac{\eta}{\tau^2} + \frac{\mu_\rho}{\sigma^2_{\rho}}\right) \sigma^2_{\rho,post}.
$$
If a truncated normal prior is used, the truncation is transferred to the full conditional distribution.

\subsection{The Full Conditional Distribution of $\boldsymbol{\beta}$}
\label{appendix:FC_beta}
%The full conditional of $\beta$ can be derived in a similar manner as the full conditional of $\alpha$. 
Analogously to (\ref{eq:general_FC}) and (\ref{eq:Likelihood}) we have 
\begin{equation}
\label{beta_FC}
p(\beta | \boldsymbol{\alpha}, \sigma^2, \boldsymbol{x}) \propto \mathcal{L}(\beta, \boldsymbol{\alpha},\sigma^2) \cdot p(\beta) = \prod_{j=0}^{t-1} \phi(x_{t-j}|\alpha_{t-j} + \beta x_{t-j-1},\sigma^2) \cdot p(\beta).
\end{equation}
By defining $\breve{d}_{t-j} := x_{t-j}-\alpha_{t-j}$ and as $$\phi(x_{t-j}|\alpha_{t-j} + \beta x_{t-j-1},\sigma^2) = \phi(\beta x_{t-j-1}|\breve{d}_{t-j},\sigma^2) 
$$
the likelihood $\mathcal{L}(\cdot)$ in the above equation can be reformulated as
\begin{align*}
\mathcal{L} &\propto \exp\left(-\frac{1}{2\sigma^2} \left\{-2\beta\left[\sum_{j=0}^{t-1}\breve{d}_{t-j} x_{t-j-1}\right]+\beta^2 \left[\sum_{j=0}^{t-1}x^2_{t-j-1}\right]\right\}\right).
\end{align*}
You can find a more detailed calculation in \ref{appendix:ReWriteLikelihood}. Defining the two terms in square brackets as $\eta$ and $\chi$, respectively, we get 
$$
\mathcal{L} \propto \exp\left(-\frac{1}{2\sigma^2} \left\{-2\beta\eta+\beta^2 \chi\right\}\right).
$$
Plugging this into (\ref{beta_FC}) and using a normal prior with parameters $\mu_{\beta}, \sigma^2_{\beta}$ for $\beta$, we have
\begin{align*}
p(\beta|\boldsymbol{\alpha},\sigma^2,\boldsymbol{x}) &\propto \exp\left(-\frac{1}{2}\left\{ \frac{\beta^2 \chi}{\sigma^2} - 2\frac{\beta\eta}{\sigma^2} \right\}\right) \exp\left(-\frac{1}{2}\left\{ \frac{\beta^2}{\sigma^2_{\beta}} - 2\frac{\beta\mu_{\beta}}{\sigma^2_{\beta}} \right\}\right)\\
&\propto \exp\left(-\frac{1}{2}\left\{\beta^2 \cdot \left(\frac{\chi}{\sigma^2} + \sigma^{-2}_{\beta} \right) -2\beta\left(\frac{\eta}{\sigma^2} + \frac{\mu_\beta}{\sigma^2_{\beta}}\right) \right\}\right)
\end{align*}
and thus $\beta | \boldsymbol{x}, \alpha, \sigma^2 \sim \mathcal{N}(\mu_{\beta,post}, \sigma^2_{\beta,post})$ with
$$\sigma^2_{\beta,post} = \left(\frac{\chi}{\sigma^2} + \sigma^{-2}_{\beta} \right)^{-1}$$ and $$\mu_{\beta,post} = \left(\frac{\eta}{\sigma^2} + \frac{\mu_\beta}{\sigma^2_{\beta}}\right) \sigma^2_{\beta,post}.
$$
If a truncated normal prior is used, the truncation is transferred to the full conditional distribution.

\subsection{The Full Conditional Distribution of $\boldsymbol{\sigma^2}$}
\label{appendix:FC_sigma2}
In this Section we derive the full conditional distribution of $\sigma^2$. As before
$$
p(\sigma^2|\boldsymbol{\alpha},\beta,\boldsymbol{x}) \propto \prod_{j=0}^{t-1} \phi(x_{t-j}|\alpha_{t-j} + \beta x_{t-j-1},\sigma^2) \cdot p(\sigma^2)\cdot p(\beta \mid \sigma^2),
$$
which is equal to 
$$
(\sigma^2)^{-\frac{t}{2}} \exp\left(-\frac{1}{2\sigma^2} \sum_{j=0}^{t-1} \epsilon^2_{t-j}\right) \cdot p(\sigma^2) \cdot p(\beta \mid \sigma^2) =: (\sigma^2)^{-\frac{t}{2}} \exp\left(-\frac{1}{2\sigma^2} \kappa \right) \cdot p(\sigma^2) \cdot p(\beta \mid \sigma^2).
$$
By using an inverse gamma distribution with shape and scale parameters $a,b$, or short $\mathcal{IG}(a,b)$, for the prior of $\sigma^2$ we get
\begin{alignat*}{1}
    &(\sigma^2)^{-(t/2)} \exp\left(-\frac{1}{2\sigma^2} \kappa \right)\cdot (\sigma^2)^{-(a+1)} \exp(-b/\sigma^2) \\
    &\qquad \qquad\cdot (\sigma^2)^{-\frac1{2}}\exp \left(-\frac1{2\sigma^2\sigma_\beta^2}(\beta-\mu_\beta)^2\right) \frac{1}{\Phi(\frac{1-\mu_\beta}{\sigma\sigma_\beta})-\Phi(\frac{-1-\mu_\beta}{\sigma\sigma_\beta})}=\\
&(\sigma^2)^{-(t/2)} \exp\left(-\frac{1}{2\sigma^2} \kappa \right)\cdot (\sigma^2)^{-\left(a+\frac3{2}\right)} \exp\left(-\frac{b+\frac{(\beta-\mu_\beta)^2}{2\sigma_\beta^2}}{\sigma^2}\right)\frac{1}{\Phi(\frac{1-\mu_\beta}{\sigma\sigma_\beta})-\Phi(\frac{-1-\mu_\beta}{\sigma\sigma_\beta})} \approx\\
&(\sigma^2)^{-\left(\frac{t+1}{2}+a + 1\right)}\exp\left(-\frac{\frac{\kappa}{2} + b + \frac{(\beta-\mu_{\beta})^2}{2\sigma_{\beta}^2}}{\sigma^2} \right).
\end{alignat*}
In the last step we omitted the last term, which results from the truncation, as in  our application the truncation is not very restrictive such that this term is close to 1. Thus the full conditional distribution is approximately also an inverse gamma distribution with parameters $\tilde{a} = \frac{t+1}{2} + a$ and $\tilde{b} = \frac{\kappa}{2} + b + \frac{(\beta-\mu_{\beta})^2}{2\sigma_{\beta}^2}$, i.e.,
$$
\sigma^2|\boldsymbol{\alpha},\beta,\boldsymbol{x} \sim \mathcal{IG}(\tilde{a},\tilde{b}).
$$

\subsection{Rewriting the Likelihood of the Parameters}
\label{appendix:ReWriteLikelihood}
By defining $\breve{d}_{t-j} := x_{t-j}-\alpha_{t-j}$, the likelihood of the parameters can be reformulated as follows: 
\begin{align*}
 \mathcal{L}(\beta,\boldsymbol{\alpha},\sigma^2) &= \prod_{j=0}^{t-1} \phi(x_{t-j}|\alpha_{t-j} + \beta x_{t-j-1},\sigma^2)\\
    &= \prod_{j=0}^{t-1} \frac{1}{\sqrt{2\pi \sigma^2}} \exp\left(-\frac{(x_{t-j}-\alpha_{t-j} - \beta x_{t-j-1})^2}{2 \sigma^2}\right)\\
    &= \prod_{j=0}^{t-1} \frac{1}{\sqrt{2\pi \sigma^2}} \exp\left(-\frac{(\breve{d}_{t-j} - \beta x_{t-j-1})^2}{2 \sigma^2}\right) \\
   & = \prod_{j=0}^{t-1} \frac{1}{\sqrt{2\pi \sigma^2}} \exp\left(-\frac{(\breve{d}_{t-j}^2 - 2\beta x_{t-j-1}\breve{d}_{t-j} +  \beta ^2 x_{t-j-1}^2}{2 \sigma^2}\right) \\
    & \propto \prod_{j=0}^{t-1} \exp\left(-\frac{1}{2 \sigma^2}\left\{ - 2\beta x_{t-j-1}\breve{d}_{t-j} +  \beta ^2 x_{t-j-1}^2\right\}\right)\\
     &= \exp\left(-\frac{1}{2 \sigma^2} \left\{ -2\beta \sum_{j=0}^{t-1}\breve{d}_{t-j} x_{t-j-1} + \beta ^2 \sum_{j=0}^{t-1}x_{t-j-1}^2\right\}\right).
\end{align*}

\section{Metropolis-Hastings within Gibbs Sampler Routine}
\label{appendix:algorithm}
Starting with an initial sample $(\alpha^{(0)}, \beta^{(0)}, (\sigma^2)^{(0)}, \rho^{(0)}, (\tau^2)^{(0)})$, where 
$$
(\tau^2)^{(0)} = f(\beta^{(0)}, (\sigma^2)^{(0)}, \rho^{(0)}, Var(x_t))
$$ 
as specified in (\ref{eq:priorTauSq}), we first draw a sample of $\boldsymbol{\widetilde{\alpha}}$ values from its full conditional distribution. We proceed with the Metropolis-Hastings algorithm step by drawing from the conditional distributions of $\rho$, $\sigma^2$ and $\beta$ as derived in Section \ref{sec:BayesianInference}. Furthermore, $\tau^2$ is set according to (\ref{eq:uncondVariance}) such that a prior specified long run variance is met. We calculate the density value of the proposal distribution $q$ specified in (\ref{eq:proposal}), i.e.,
\begin{align*}
q(\rho^{(t+1)}, &\beta^{(t+1)}, (\sigma^2)^{(t+1)} \mid (\tau^2)^{(t)}, \beta^{(t)}, (\sigma^2)^{(t)}, \boldsymbol{\alpha}, \boldsymbol{x}) = \\ &q(\rho^{(t+1)} \mid (\tau^2)^{(t)}, \boldsymbol{\alpha}, x) q((\sigma^2)^{(t+1)} \mid \beta^{(t)} \boldsymbol{\alpha}, \boldsymbol{x}) q(\beta^{(t+1)} \mid (\sigma^2)^{(t)}, \boldsymbol{\alpha}, \boldsymbol{x})
\end{align*}
We further calculate the density value of the proposal distribution for the parameters of the previous step conditional on the new drawn parameter, i.e.,
\begin{align*}
q(\rho^{(t)}, &\beta^{(t)}, (\sigma^2)^{(t)} \mid (\tau^2)^{(t+1)}, \beta^{(t+1)}, (\sigma^2)^{(t+1)}, \boldsymbol{\alpha}, x) = \\ &q(\rho^{(t)} \mid (\tau^2)^{(t+1)}, \boldsymbol{\alpha}, x) q((\sigma^2)^{(t)} \mid \beta^{(t+1)} \boldsymbol{\alpha}, x) q(\beta^{(t)} \mid (\sigma^2)^{(t+1)}, \boldsymbol{\alpha}, x) 
\end{align*}
The true conditional posterior density is given by 
\begin{align*}
    p(\rho, \beta, \sigma^2 \mid \boldsymbol{\alpha}, \boldsymbol{x}) \propto p(\boldsymbol{x} \mid \beta, \sigma^2, \boldsymbol{\alpha}) p(\boldsymbol{\alpha} \mid \rho, f(\rho, \beta, \sigma^2))p(\rho)p(\beta \mid \sigma^2) p(\rho)
\end{align*}
The acceptance probability is calculated by
$$
p_{accept.} = \min \left( 1, \frac{p(\rho^{(t+1)}, \beta{(t+1)}, (\sigma^2)^{(t+1)} \mid \boldsymbol{\alpha^{(t+1)}}, \boldsymbol{x})q(\rho^{(t)}, \beta^{(t)}, (\sigma^2)^{(t)} \mid \boldsymbol{\alpha^{(t+1)}}, \boldsymbol{x})}{p(\rho^{(t)}, \beta{(t)}, (\sigma^2)^{(t)} \mid \boldsymbol{\alpha^{(t+1)}}, \boldsymbol{x}) q(\rho^{(t+1)}, \beta^{(t+1)}, (\sigma^2)^{(t+1)} \mid \boldsymbol{\alpha^{(t+1)}}, \boldsymbol{x})}\right)
$$
A new drawn sample is accepted if a uniform distributed random variable is smaller than the acceptance probability. Otherwise the sample from the previous step is taken. After a burn-in period the parameter set $(\boldsymbol{\widetilde{\alpha}}^{(m)}, \beta^{(m)}, (\sigma^2)^{(m)})$ is approximately distributed according to the joint posterior distribution $p(\boldsymbol{\widetilde{\alpha}}, \beta, \sigma^2 \mid \boldsymbol{x})$.

%\begin{algorithm}[H]
%\caption{}
%\begin{algorithmic}[1]
%  \scriptsize
%  \STATE Simulate from the full conditional distribution of %$\boldsymbol{\widetilde{\alpha}}$ by using the latest sample %of the other parameters:
%  \begin{itemize}
%      \item $\boldsymbol{\widetilde{\alpha}^{(m+1)}} \sim p(\boldsymbol{\widetilde{\alpha}} \mid \rho^{(m)}, (\tau^2)^{(m)}, \beta^{(m)}, (\sigma^2)^{(m)}, x)$
%  \end{itemize}
%  \STATE Simulate from the full conditional distribution of $\rho$, the hyper parameter of $\boldsymbol{\widetilde{\alpha}}$, by using the latest sample of $\boldsymbol{\alpha}$ and $\tau^2$:
%  \begin{itemize}
%      \item $\rho^{(m+1)} \sim p(\rho \mid \boldsymbol{\alpha^{(m+1)}}, (\tau^2)^{(m)}, x)$
%  \end{itemize}
%  \STATE Simulate from the full conditional distribution of $\sigma^2$ by using the latest sample of $\boldsymbol{\alpha}$ and $\beta^2$:
%  \begin{itemize}
%      \item $(\sigma^2)^{(m+1)} \sim p((\sigma^2)^{(m+1)} \mid \boldsymbol{\alpha}^{(m+1)}, \beta^{(m)}, x)$
%  \end{itemize}
%  \STATE Simulate from the full conditional distribution of $\beta|\sigma^2$ by using the latest sample of $\boldsymbol{\alpha}$ and $\sigma^2$:
%  \begin{itemize}
%      \item $\beta^{(m+1)} \sim p(\beta \mid \boldsymbol{\alpha}^{(m+1)}, (\sigma^2)^{(m+1)}, x)$
%  \end{itemize}
%  \STATE Set $(\tau^2)^{(m+1)}$ according to (\ref{eq:uncondVariance}) such that a prior specified long run variance is met.
%\end{algorithmic}
%\end{algorithm}

\section{Backtest Results}
\label{appendix:Tables}
\begin{table} [H]
\begin{tabularx}{\textwidth}{p{4.5cm}p{3.5cm}p{3.5cm}p{3.5cm}}
 Maturity & Mean & Std. Dev. & RMSE\\
 \hline
 \hline
 \\
 \multicolumn{4}{l}{\textit{The BTVC-AR(1)-Factor model}} \\
 1 year & -0.0268 & 0.2566 & 0.0659  \\
 3 year & -0.0469 & 0.2289 & 0.0541  \\
 5 year & -0.0681 & 0.2402 & 0.0617  \\
 10 year & -0.0640 & 0.2346 & 0.0586 \\
 \\
 \multicolumn{4}{l}{\textit{The Gauss2++ model}} \\
 1 year & -0.0808 & 0.2361 & 0.0618  \\
 3 year & -0.1037 & 0.2252 & 0.0610  \\
 5 year & -0.1203 & 0.2139 & 0.0598  \\
 10 year & -0.1429 & 0.2130 & 0.0654 \\
 \\
  \multicolumn{4}{l}{\textit{The dynamic Nelson-Siegel model}} \\
 1 year & -0.0290 & 0.2615 & 0.0685  \\
 3 year & -0.0462 & 0.2311 & 0.0550  \\
 5 year & -0.0653 & 0.2410 & 0.0617  \\
 10 year & -0.0589 & 0.2340 & 0.0577 \\
 \hline
\end{tabularx}
\caption{Results of the out-of-sample 1-month ahead forecasting.}
 \label{tbl:1monthFC}
\end{table}
\begin{table} [H]
\begin{tabularx}{\textwidth}{p{4.5cm}p{3.5cm}p{3.5cm}p{3.5cm}}
 Maturity & Mean & Std. Dev. & RMSE\\
 \hline
 \hline
 \\
 \multicolumn{4}{l}{\textit{The BTVC-AR(1)-Factor model}} \\
 1 year & -0.1264 & 0.5064 & 0.2697 \\
 3 year & -0.1505 & 0.4640 & 0.2358  \\
 5 year & -0.1725 & 0.4354 & 0.2174 \\
 10 year & -0.1625 & 0.3875 & 0.1751 \\
 \\
 \multicolumn{4}{l}{\textit{The Gauss2++ model}} \\
 1 year & -0.2057 & 0.5329 & 0.3236  \\
 3 year & -0.2707 & 0.4702 & 0.2923  \\
 5 year & -0.3098 & 0.4208 & 0.2714  \\
 10 year & -0.3435 & 0.3875 & 0.2667 \\
 \\
  \multicolumn{4}{l}{\textit{The dynamic Nelson-Siegel model}} \\
 1 year & -0.1327 & 0.5152 & 0.2803  \\
 3 year & -0.1482 & 0.4665 & 0.2374  \\
 5 year & -0.1643 & 0.4343 & 0.2137  \\
 10 year & -0.1478 & 0.3827 & 0.1668 \\
 \hline
\end{tabularx}
\caption{Results of the out-of-sample 3-month ahead forecasting.}
 \label{tbl:3monthFC}
\end{table}
\vspace{0.25cm}
\begin{table} [H]
\begin{tabularx}{\textwidth}{p{4.5cm}p{3.5cm}p{3.5cm}p{3.5cm}}
 Maturity & Mean & Std. Dev. & RMSE\\
 \hline
 \hline
 \\
 \multicolumn{4}{l}{\textit{The BTVC-AR(1)-Factor model}} \\
 1 year & -0.2809 & 0.7683 & 0.6631  \\
 3 year & -0.3093 & 0.6941 & 0.5725  \\
 5 year & -0.3311 & 0.6330 & 0.5062  \\
 10 year & -0.3110 & 0.5462 & 0.3920 \\
 \\
 \multicolumn{4}{l}{\textit{The Gauss2++ model}} \\
 1 year & -0.4094 & 0.8105 & 0.8184  \\
 3 year & -0.5402 & 0.6768 & 0.7457  \\
 5 year & -0.6098 & 0.6090 & 0.7393  \\
 10 year & -0.6545 & 0.5824 & 0.7693 \\
 \\
  \multicolumn{4}{l}{\textit{The dynamic Nelson-Siegel model}} \\
 1 year & -0.2900 & 0.7857 & 0.6951  \\
 3 year & -0.3022 & 0.7045 & 0.5825  \\
 5 year & -0.3130 & 0.6380 & 0.5008  \\
 10 year & -0.2812 & 0.5446 & 0.3727 \\
 \hline
\end{tabularx}
\caption{Results of the out-of-sample 6-month ahead forecasting.}
 \label{tbl:6monthFC}
\end{table}
\vspace{0.25cm}
\begin{table} [H]
\begin{tabularx}{\textwidth}{p{4.5cm}p{3.5cm}p{3.5cm}p{3.5cm}}
 Maturity & Mean & Std. Dev. & RMSE\\
 \hline
 \hline
 \\
 \multicolumn{4}{l}{\textit{The BTVC-AR(1)-Factor model}} \\
 1 year & -0.5956 & 0.9591 & 1.2652  \\
 3 year & -0.6264 & 0.7861 & 1.0041  \\
 5 year & -0.6526 & 0.6834 & 0.8881  \\
 10 year & -0.6275 & 0.5986 & 0.7484 \\
 \\
 \multicolumn{4}{l}{\textit{The Gauss2++ model}} \\
 1 year & -0.9047 & 1.0709 & 1.9546  \\
 3 year & -1.1531 & 0.7939 & 1,9541  \\
 5 year & -1.2745 & 0.7255 & 2.1458  \\
 10 year & -1.3345 & 0.8060 & 2.4246 \\
 \\
  \multicolumn{4}{l}{\textit{The dynamic Nelson-Siegel model}} \\
 1 year & -0.6004 & 0.9961 & 1.3424  \\
 3 year & -0.6024 & 0.8218 & 1.0316  \\
 5 year & -0.6098 & 0.7096 & 0.8702  \\
 10 year & -0.5657 & 0.6024 & 0.6793 \\
 \hline
\end{tabularx}
\caption{Results of the out-of-sample 12-month ahead forecasting.}
 \label{tbl:12monthFC}
\end{table}

\end{document}